\documentclass[a4paper]{article}%
\usepackage{amsfonts}
\usepackage{amsmath}
\usepackage{amssymb}
\usepackage{amstext}
\usepackage{graphicx}
\usepackage{appendix}%
\providecommand{\U}[1]{\protect\rule{.1in}{.1in}}
\newtheorem{theorem}{Theorem}[section]

\newtheorem{lemma}[theorem]{Lemma}

\newtheorem{proposition}[theorem]{Proposition}
\newtheorem{remark}[theorem]{Remark}

\newenvironment{preuve}[1][Proof]{\noindent\textbf{#1.} }{\
  \rule{0.5em}{0.5em}}
\numberwithin{equation}{section}
\begin{document}

\title{An explicit model for the adiabatic evolution of quantum observables driven by
1D shape resonances.}
\author{A. Faraj$^{\ast}$, A. Mantile$^{\ast}$, F. Nier\thanks{IRMAR, UMR - CNRS 6625,
Universit\'{e} Rennes 1, Campus de Beaulieu, 35042 Rennes Cedex, France.}}
\date{\emph{Dedicated to the memory of P. Duclos.}}
\maketitle

\begin{abstract}
This paper is concerned with a linearized version of the quantum transport
problem where the Schr\"{o}dinger-Poisson operator is replaced by a
non-autonomous Hamiltonian, slowly varying in time. We consider an explicitly
solvable system where a semiclassical island is described by a flat potential
barrier, while a time dependent 'delta' interaction is used as a model for a
single quantum well. Introducing, in addition to the complex deformation, a
further modification formed by artificial interface conditions, we give a
reduced equation for the adiabatic evolution of the sheet density of charges
accumulating around the interaction point.

\end{abstract}

\section{Introduction}

The derivation of reduced models for the dynamics of transverse quantum
transport with concentrated non-linearities plays a central role in the
mathematical analysis of semiconductor heterostructures like tunneling diodes
or possibly more complex structures. The conduction band edge-profile of such
systems has been described using Schr\"{o}dinger-Poisson Hamiltonians with
quantum wells in a semiclassical island, where a non-linear potential term,
depending on the local charge density, approximates in the mean field limit
the repulsive interaction between the charge carriers. A functional framework
for such a model is proposed in \cite{Ni1}, based on Mourre's theory and
Sigal-Soffer propagation estimates, and implements a dynamical nonlinear
version of the Landauer-B\"{u}ttiker approach. The analysis of the related
steady state problem, developed in \cite{BNP1}, \cite{BNP2}, \cite{Ni2} on the
basis of the Helffer-Sj\"{o}strand approach to resonances \cite{HeSj1}, has
provided with an asymptotic reduced equation for the nonlinear potential,
which elucidates the influence of the geometry of the potential on the
feasibility of hysteresis phenomena, already studied in \cite{JoPrSj},
\cite{PrSj}, and confirms the general belief arising in physical literature:
The nonlinear phenomena are governed by a finite number of resonant states.

For the dynamical problem, we conjecture that the nonlinear dynamics follows
the time evolution of those resonant states corresponding to shape resonances
which are asymptotically embedded in some relevant energy interval when the
quantum scale of the problem, parametrized by $h$, goes to zero. It is known,
at least in the linear case, that this evolution shows an exponential decay
behaviour having physical interpretation in terms of truncated resonant states
(lying in $L^{2}$). The \emph{quasi-resonant states }concentrate their mass
inside the quantum well's support -- the classical region of motion of our
model -- on a long time scale given by the inverse of the imaginary part of
the resonant energies $E_{res}^{h}$. In this connection, the Poisson
potential, as well as the charge density for the nonlinear modelling, are
expected to evolve slowly in time, with an adiabatic parameter $\varepsilon$
which is related to the quantum scale of the system according to:
$\varepsilon=\mathcal{O}(\operatorname{Im}E_{res}^{h})\sim e^{-\frac{\tau}{h}%
}$, for some $\tau>0$.

This paper is concerned with a linearized version of the transport problem
where the Schr\"{o}dinger-Poisson operator is replaced by a non-autonomous
Hamiltonian, slowly varying in time, and whose time profile takes into account
the evolution of the nonlinear potential. This allows us to separate the
adiabatic evolution generated by the double scale Hamiltonian, from questions
concerned with the nonlinear nature of the original problem. In particular, we
consider an explicitly solvable model where the semiclassical island is
described by a flat potential barrier, while a time dependent 'delta'
interaction describes, with the suitable scaling, a single quantum well. Our
approach consists in introducing, in addition to the complex deformation, a
further modification formed by artificial interface conditions. According to
the results obtained in \cite{FMN2}, an adiabatic theorem holds for this
modified system (see Theorem~7.1 in \cite{FMN2}), while small perturbations
are produced on the relevant spectral quantities (actually the same remains
true under more general assumptions). In this simplified framework, we give a
reduced equation for the adiabatic evolution of the sheet density of charges
accumulating around the interaction point. This result is coherent with the
reduced model predicted in \cite{PrSj},\cite{PrSj1}. Moreover, some
corrections arise, depending on the time profile of the perturbation, which
can be relevant in realistic physical situations.

\section{\label{Sec_model}The model}

We consider the time evolution of a quantum observable for a family of
non-selfadjoint Hamiltonians adiabatically depending on the time. Our model is
defined by the Schr\"{o}dinger operators $H_{\theta_{0},\alpha(t)}^{h}$%
\begin{equation}
H_{\theta_{0},\alpha(t)}^{h}=-h^{2}\Delta_{\theta_{0}}+1_{\left(  a,b\right)
}V_{0}+h\alpha(t)\delta_{c}\,, \label{H}%
\end{equation}
where $\Delta_{\theta_{0}}$ is a singularly perturbed Laplacian with
artificial interface conditions on the boundary of $\mathbb{R}\backslash
\left\{  a,b\right\}  $%
\begin{equation}
\left\{
\begin{array}
[c]{l}%
\medskip D(\Delta_{\theta_{0}})=\left\{  u\in H^{2}(\mathbb{R}\backslash
\left\{  a,b\right\}  ):\left[
\begin{array}
[c]{l}%
\smallskip e^{-\frac{\theta_{0}}{2}}u(b^{+})=u(b^{-});\ e^{-\frac{3}{2}%
\theta_{0}}u^{\prime}(b^{+})=u^{\prime}(b^{-})\\
e^{-\frac{\theta_{0}}{2}}u(a^{-})=u(a^{+});\ e^{-\frac{3}{2}\theta_{0}%
}u^{\prime}(a^{-})=u^{\prime}(a^{+})
\end{array}
\right.  \right\} \\
\Delta_{\theta_{0}}u=\partial_{x}^{2}u\,.
\end{array}
\right.  \,, \label{Laplacian_mod}%
\end{equation}
Meanwhile $1_{\left(  a,b\right)  }V_{0}+h\alpha\delta_{c}$ is a selfadjoint
time dependent point interaction defined with: $V_{0}>0$, $c\in\left(
a,b\right)  $, $\alpha\in C^{\infty}(0,T)$ and requiring the condition%
\begin{equation}
u\in H^{2}(\left(  a,b\right)  \backslash\left\{  c\right\}  )\cap
H^{1}(a,b)\,,\qquad h\left[  u^{\prime}(c^{+})-u^{\prime}(c^{-})\right]
=\alpha(t)u(c), \label{B.C.delta}%
\end{equation}
for all $u\in D(H_{\theta_{0},\alpha(t)}^{h})$ (we refer to \cite{Albeverio}
for the definition of delta interaction Hamiltonians).

An accurate analysis of this class of operators has been given, in
\cite{FMN2}. It is shown that the interface conditions introduce small errors,
controlled by $\theta_{0}$, with respect to the original selfadjoint model
(defined by $\theta_{0}=0$). The main interest in introducing the artificial
perturbation $\Delta_{\theta_{0}}$ rests upon the fact that the corresponding
Hamiltonian defines, under complex deformation, a dynamical systems of
contractions. This provides us with an alternative approach to the adiabatic
evolution of the shape resonances possibly associated with our model, which
can be treated in terms of (adiabatic evolution of) spectral projectors for
the non-selfadjoint deformed operator (for this point, we refer to Theorem~7.1
in \cite{FMN2}; see also the work of A. Joye \cite{Joye1} for the adiabatic
evolution of dynamical systems without uniform time estimates on the semigroup).

Let us consider a positive smooth function $\chi$, supp $\chi\subset(a,b)$; in
our framework $\chi$ is the quantum observable associated with the charge
density accumulated in a small neighbourhood of the quantum well. The expected
value of this density sheet is associated with%
\begin{equation}
A_{\theta_{0}}(t)=Tr\left[  \chi\,\rho_{t}^{h}\right]  \,, \label{A}%
\end{equation}
where $\rho_{t}^{h}$ is the time evolution of the density operator. The
initial state of the system,%
\begin{equation}
\rho_{0}^{h}=\int\frac{dk}{2\pi h}g(k)\,\left\vert \psi_{-}(k,\cdot,\alpha
_{0})\right\rangle \,\left\langle \psi_{-}(k,\cdot,\alpha_{0})\right\vert \,,
\label{initial_state}%
\end{equation}
is defined by a superposition of incoming scattering states solving%
\[
\left(  H_{\theta_{0},\alpha}^{h}-k^{2}\right)  \psi_{-}(k,\cdot,\alpha)=0\,,
\]
according to the out-of-equilibrium assumption $g=1_{\mathbb{R}_{+}}g$. Using
an adiabatic approximation for the time variations of the coupling parameter
$\alpha$, $\rho_{t}^{h}$ writes as%
\begin{equation}
\rho_{t}^{h}=\int\frac{dk}{2\pi h}g(k)\,\left\vert u(k,\cdot,t)\right\rangle
\,\left\langle u(k,\cdot,t)\right\vert \,, \label{density}%
\end{equation}
with%
\begin{equation}
\left\{
\begin{array}
[c]{l}%
\medskip i\varepsilon\partial_{t}u(k,\cdot,t)=H_{\theta_{0},\alpha(t)}%
^{h}u(k,\cdot,t)\,,\\
u_{t=0}=\psi_{-}(k,\cdot,\alpha_{0})\,.
\end{array}
\right.  \label{evolution}%
\end{equation}

Adiabatic dynamics have already been considered within the modelling of
out-of-equilibrium quantum transport in \cite{AEGS}, \cite{AEGSS},
\cite{CDNP}, playing with the continuous spectrum with selfadjoint techniques.
For energies close to the shape resonances, the relevant observable of this
problem follow the adiabatic evolution of resonant states. Then, a different
approach consists in using complex deformations, originally introduced in
\cite{AgCo}, \cite{BaCo}. In \cite{FMN2}, we define a family of exterior
complex deformations $U_{\theta}$ for Hamiltonians with compactly supported
potentials in $(a,b)$%
\begin{equation}
U_{\theta}u(x)=\left\{
\begin{array}
[c]{l}%
e^{\frac{\theta}{2}}u(\smallskip e^{\theta}(x-b)+b),\qquad x>b\,,\\
\smallskip u(x),\qquad\qquad\qquad\qquad\ x\in(a,b)\,,\\
e^{\frac{\theta}{2}}u(\smallskip e^{\theta}(x-a)+a),\qquad x<a\,.
\end{array}
\right.  \label{U_theta}%
\end{equation}
The corresponding deformed operator is obtained by conjugation: $H_{\theta
_{0},\alpha(t)}^{h}(\theta)=U_{\theta}H_{\theta_{0},\alpha(t)}^{h}U_{\theta
}^{-1}$\,. It is explicitly written as%
\begin{equation}
H_{\theta_{0},\alpha(t)}^{h}(\theta)=-h^{2}e^{-2\theta\,1_{\mathbb{R}%
\backslash\left(  a,b\right)  }}\Delta_{\theta_{0}+\theta}+1_{\left(
a,b\right)  }V_{0}+h\alpha(t)\delta_{c}\,. \label{H_def}%
\end{equation}

Since $\chi$ commutes with $U_{\theta}$ for all values $\theta\in\mathbb{C}$,
the variable $A_{\theta_{0}}(t)$ can be defined in terms of deformed
quantities. Thus $A_{\theta_{0}}(t)$ can be rephrased as%
\begin{equation}
A_{\theta_{0}}(t)=Tr\left[  U_{\theta}^{\ast}\chi U_{\theta}\rho_{t}%
^{h}\right]  =Tr\left[  \chi U_{\theta}\rho_{t}^{h}U_{\theta}^{\ast}\right]
\,.
\end{equation}
Denoting with $S_{\theta_{0},\varepsilon}(t,s)$ the time propagator related to
$\frac{1}{\varepsilon}H_{\theta_{0},\alpha(t)}^{h}$, we get%
\[
A_{\theta_{0}}(t)=Tr\left[  \chi U_{\theta}S_{\theta_{0},\varepsilon}%
(t,0)\rho_{0}^{h}S_{\theta_{0},\varepsilon}^{\ast}(t,0)U_{\theta}^{\ast
}\right]  =Tr\left[  \chi U_{\theta}S_{\theta_{0},\varepsilon}(t,0)U_{\theta
}^{-1}U_{\theta}\rho_{0}^{h}U_{\theta}^{\ast}\left(  U_{\theta}^{-1}\right)
^{\ast}S_{\theta_{0},\varepsilon}^{\ast}(t,0)U_{\theta}^{\ast}\right]  \,,
\]
where $\left(  U_{\theta}^{-1}\right)  ^{\ast}=\left(  U_{\theta}^{\ast
}\right)  ^{-1}$ is used. The conjugation: $U_{\theta}S_{\theta_{0}%
,\varepsilon}(t,0)U_{\theta}^{-1}$ defines the propagator associated with the
deformed Hamiltonian $\frac{1}{\varepsilon}H_{\theta_{0},\alpha(t)}^{h}%
(\theta)$. Thus (\ref{A}) reformulates as follows%
\begin{equation}
A_{\theta_{0}}(t)=Tr\left[  \chi\rho_{t}^{h}(\theta)\right]  \,, \label{A_def}%
\end{equation}%
\begin{equation}
\rho_{t}^{h}(\theta)=\int\frac{dk}{2\pi h}g(k)\,\left\vert u_{\theta}%
(k,\cdot,t)\right\rangle \,\left\langle u_{\theta}(k,\cdot,t)\right\vert \,,
\label{A_def1}%
\end{equation}
with%
\begin{equation}
\left\{
\begin{array}
[c]{l}%
\medskip i\varepsilon\partial_{t}u_{\theta}(k,\cdot,t)=H_{\theta_{0}%
,\alpha(t)}^{h}(\theta)u_{\theta}(k,\cdot,t)\\
u_{t=0}=U_{\theta}\psi_{-}(k,\cdot,\alpha_{0})
\end{array}
\right.  \,. \label{A_def2}%
\end{equation}

We will consider this evolution problem under the following assumptions:

\begin{description}
\item[h1)] The deformation and the interface conditions parameters are equals
and%
\begin{equation}
\theta=\theta_{0}=ih^{N_{0}},\qquad N_{0}>2 \label{h1}%
\end{equation}

\item[\textbf{h2)}] The time dependent coupling parameter $\alpha_{t}$ is a
$C^{\infty}(0,T)$ real valued function with compact range in $\left(
-2V_{0}^{\frac{1}{2}},0\right)  $ and such that:\newline$i)$ Its firsts
variations have size $h$, i.e.:%
\begin{equation}
\forall s,t\in\left[  0,T\right]  \Rightarrow\left\vert \alpha_{t}-\alpha
_{s}\right\vert \leq\frac{2h}{\sqrt{V_{0}}\,d_{0}} \label{alpha_var}%
\end{equation}
where $d_{0}>0$ is specified further. \newline$ii)$ There exists a positive
integer $J$ such that the vector $\left\{  \partial_{t}^{j}\alpha(t)\right\}
_{j=1}^{J}$ is not null for all $t$.\newline

\item[h3)] The initial state is defined with a smooth and compactly supported
partition function $g$ such that:%
\begin{equation}
\text{supp }g(k)=\left\{  k>0,\ \left\vert k^{2}-\lambda_{0}\right\vert
<2\frac{h}{d_{0}}\right\}  \label{h2}%
\end{equation}
where $\lambda_{0}$ denotes some asymptotic energy: $\lambda_{0}\in\left(
0,V_{0}\right)  $, while $d_{0}$ and $h_{0}$ are such that: supp
$g\subset\left(  0,V_{0}\right)  $ uniformly w.r.t. $h\in(0,h_{0})$\,.
\newline Furthermore, we assume that $g(E^{\frac{1}{2}})$ extends to an
holomorphic function of $E$ in the complex neighbourhood of $\lambda_{0}$ of
radius $\frac{h}{d_{0}}$. \newline The function $\chi\in C_{0}^{\infty}(a,b)$
is real valued and such that:%
\begin{equation}
\left\{
\begin{array}
[c]{l}%
\medskip\chi=\left(  c-2\eta,c+2\eta\right) \\
\left.  \chi(x)\right\vert _{x\in\left(  c-\eta,c+\eta\right)  }=1
\end{array}
\right.  ,\qquad\eta<d(c,\left\{  a,b\right\}  )
\end{equation}
$d(c,\left\{  a,b\right\}  )$ denoting the distance of $c$ from the the
boundary of the interval $\left(  a,b\right)  $.

\item[h4)] The adiabatic parameter is fixed to the exponential scale defined
by%
\begin{equation}
\varepsilon=e^{-\frac{\left\vert \alpha_{0}\right\vert }{h}\,d(c,\left\{
a,b\right\}  )}\,, \label{eps}%
\end{equation}

\end{description}

The explicit character of our model and the adiabatic theorem, obtained in
\cite{FMN2} for this class of non-selfadjoint Hamiltonians, allow to obtain a
complete description of the asymptotic behaviour of $A_{\theta_{0}}(t)$ as
$h\rightarrow0$, in particular concerned with the position of the delta shaped
potential well. To formulate our results, we adopt the following notation.

\begin{description}
\item[Notation] a) The resonance at time $t$ and the related resonant state
are respectively denoted with $E(t)$ and $G(t)$.\newline b) The expression:
$X_{\varepsilon}=\mathcal{\tilde{O}}(\varepsilon^{n})$ is used for the
following condition: $\forall$ $\delta\in(0,1)$, there exists $C_{X,\delta}$
such that%
\begin{equation}
\left\vert X_{\varepsilon}\right\vert \leq C_{X,\delta}\,\varepsilon
^{n-\delta}\,. \label{Otilda}%
\end{equation}

\end{description}

\begin{theorem}
\label{Th_1}Let $\lambda_{t}$ $=V_{0}-\frac{\alpha_{t}^{2}}{4}\,$,
$t\in\left[  0,T\right]  $ and assume the conditions (h1)-(h4) to hold with:
$h\in\left(  0,h_{0}\right)  $, $h_{0}$ small, $\lambda_{0}=V_{0}-\frac
{\alpha_{0}^{2}}{4}$, and $d>0$ such that: $\lambda_{t}^{\frac{1}{2}}\in$ supp
$g\subset\left(  0,V_{0}\right)  $ for all $t$. The following conditions
hold:\medskip\newline$i)$ For any $t\in\left[  0,T\right]  $, there exists a
single resonance, $E(t)$, of $H_{\theta_{0},\alpha(t)}^{h}$ such that:
$\operatorname{Re}E^{\frac{1}{2}}(t)\in\left(  0,V_{0}\right)  $. With the
notation: $E(t)=E_{R}(t)-i\Gamma_{t}\,$, the real and the imaginary parts of
$E(t)$ fulfill the conditions%
\begin{equation}
E_{R}(t)=\lambda_{t}+\mathcal{O}\left(  e^{-\frac{\left\vert \alpha
_{t}\right\vert }{h}\,d(c,\left\{  a,b\right\}  )}\right)  \,,\label{h2_real}%
\end{equation}%
\begin{equation}
\Gamma_{t}=\mathcal{O}\left(  e^{-\frac{\left\vert \alpha_{t}\right\vert }%
{h}\,d(c,\left\{  a,b\right\}  )}\right)  \,,\label{h2_immaginary}%
\end{equation}
$d(\cdot,\left\{  a,b\right\}  )$ denoting the distance from the boundary
points. The related resonant state, $G(t)$, is locally defined as the solution
of%
\begin{equation}
\left(  H_{\theta_{0},0}^{h}-E(t)\right)  u=\delta_{c}\,,\qquad\text{in }%
L^{2}(a,b)\,.
\end{equation}
Both $E(t)$ and $G(t)$ are holomorphic w.r.t. $\alpha$, and $C^{\infty}$ in
time. \medskip\newline$ii)$ There exists $\tau_{\chi,J}>0$, depending on
$\chi,J$, such that the solution of (\ref{A_def})-(\ref{A_def2}) is%
\begin{equation}
A_{\theta_{0}}(t)=a(t)+\mathcal{J}(t)+\mathcal{O}\left(  \left\vert \theta
_{0}\right\vert \right)  +\mathcal{\tilde{O}}\left(  e^{-\frac{\tau_{\chi,J}%
}{h}}\right)  \,.
\end{equation}
The main contribution, $a(t)$, is described by the equation%
\begin{equation}
\left\{
\begin{array}
[c]{l}%
\partial_{t}a(t)=\left(  -2\frac{\Gamma_{t}}{\varepsilon}\right)  \,\left(
a(t)-\left\vert \frac{\alpha_{t}}{\alpha_{0}}\right\vert ^{3}g\left(
\lambda_{t}^{\frac{1}{2}}\right)  \right)  \\
a(0)=g\left(  \lambda_{0}^{\frac{1}{2}}\right)  \,
\end{array}
\right.  ,\qquad\text{for }d(c,\left\{  a,b\right\}  )=c-a\,,
\end{equation}
or by: $a(t)=$ $\mathcal{O}\left(  e^{-\frac{\beta}{h}}\right)  $,
$\beta=\frac{\left\vert \alpha_{t}\right\vert }{h}(c-a-(b-c))$ if
$d(c,\left\{  a,b\right\}  )=b-c$.\newline$iii)$ When $d(c,\left\{
a,b\right\}  )=c-a$, the remainder is: $\mathcal{J}=\mathcal{J}_{1}%
+\mathcal{J}_{2}+\mathcal{\tilde{O}}\left(  e^{-\frac{\tau_{\chi,K}}{h}%
}\right)  $\thinspace,%
\begin{equation}
\mathcal{J}_{1}(t)=\left\vert 1-\left\vert \frac{\alpha_{t}}{\alpha_{0}%
}\right\vert ^{\frac{3}{2}}\right\vert ^{2}\,g\left(  \lambda_{t}^{\frac{1}%
{2}}\right)  =\mathcal{O}\left(  h^{2}\right)  \,,
\end{equation}
while $\mathcal{J}_{2}$ generates a boundary layer contribution depending on
the difference $\lambda_{t}-\lambda_{0}$, and whose explicit form is given by%
\begin{equation}
\mathcal{J}_{2}(t)=\operatorname{Re}2i\left(  1-\left\vert \frac{\alpha_{t}%
}{\alpha_{0}}\right\vert ^{\frac{3}{2}}\right)  \,\frac{\Gamma_{t}%
}{\varepsilon}\,g\left(  \lambda_{t}^{\frac{1}{2}}\right)  \frac
{\mathcal{T}(t)}{\frac{\lambda_{t}-\lambda_{0}}{\varepsilon}-i\frac{\left(
\Gamma_{t}+\Gamma_{0}\right)  }{\varepsilon}}\,,
\end{equation}%
\begin{equation}
\mathcal{T}(t)=\frac{\left\vert \alpha_{0}\right\vert \alpha_{t}^{2}%
+\alpha_{0}^{2}\left\vert \alpha_{t}\right\vert }{\left(  \alpha_{0}\alpha
_{t}\right)  ^{\frac{3}{2}}}\,e^{-\frac{1}{\varepsilon}\int_{0}^{t}\left(
\Gamma_{\sigma}+\Gamma_{t}\right)  \,d\sigma}e^{-\frac{i}{\varepsilon}\int
_{0}^{t}\left(  \lambda_{\sigma}-\lambda_{t}\right)  \,d\sigma}\,.
\end{equation}
For $d(c,\left\{  a,b\right\}  )=b-c$, the correction $\mathcal{J}=$
$\mathcal{O}\left(  e^{-\frac{\beta}{h}}\right)  $ is exponentially small.
\end{theorem}

The above result is concerned with situations where the two barriers composing
our potential have different opacity w.r.t. the electron tunneling. When the
interaction point '$c$' is closer to the left boundary of the barrier, i.e.
$d(c,\left\{  a,b\right\}  )=c-a$, a macroscopic variation of the charges
accumulating around $c$ is observed and a reduced equation is given. This
corresponds to the appearance of macroscopic hysteresis phenomena in the
nonlinear modelling where a similar simplified equation was predicted
\cite{PrSj}. On the opposite, for $d(c,\left\{  a,b\right\}  )=b-c$, only
exponentially small contributions to $A_{\theta_{0}}$ appears as
$h\rightarrow0$. Although the critical case, given by: $b-c=c-a+\mathcal{O}%
\left(  h\right)  $, is not explicitly considered here, most of the
computations developed in this work can be adapted to study this particular problem.

The reduced model of Theorem~\ref{Th_1} follows form explicit computations,
which are made possible by our simplified setting. The coefficient $\left\vert
\frac{\alpha_{t}}{\alpha_{0}}\right\vert ^{3}$, appearing in this formulation,
arises from the ratio of the $L^{2}$ square norms of resonant functions:%
\[
\frac{\left\Vert G(0)\right\Vert _{L^{2}(\mathbb{R})}^{2}}{\left\Vert
G(t)\right\Vert _{L^{2}(\mathbb{R})}^{2}}\,.
\]
This provides a possible 'link' to extend the analysis to more realistic
situations. The remainders have size: $\mathcal{J}_{1}=\mathcal{O}\left(
h^{2}\right)  $, while the second term can be relevant whenever $\lambda
_{t}-\lambda_{0}\sim\mathcal{O}(\varepsilon)$. Under this particular condition
one has: $\mathcal{J}_{2}(t)=\mathcal{O}\left(  h\right)  $. It is pointed out
in Section \ref{Sec_proof} that $\mathcal{J}_{2}$ coincides, out of
exponentially small terms, with a scalar product between the initial state and
its evolution at time $t$. Since we are close to the resonance, the adiabatic
theorem applies, and this evolution "follows" the resonant state at time $t$.
Thus $\mathcal{J}_{2}(t)$ shows a local maximum when $E(t)\sim E(0)$ and the
corresponding resonant states are highly correlated.

Some of the assumptions in (h1)-(h4) can be relaxed according to the following
points:\newline$I)$ If we limit to the point $ii)$ of the theorem, the
condition $\alpha_{t}\in C^{2}(0,T)$ is sufficient for the derivation of the
reduced model, once that a suitable adaptation of the proof of Lemma
\ref{Lemma_diag} is provided. Nevertheless, the conditions $\alpha_{t}\in
C^{\infty}$ and (h2)-$ii)$ play a central role when the asymptotics of the
remainder terms is considered.\newline$II)$ The relation (\ref{h2}) fixes the
out-of-equilibrium condition $k>0$ for the density matrix. The
constrain$\ \left\vert k^{2}-\lambda_{0}\right\vert <2\frac{h}{d_{0}}$ selects
the leading term of the density kernel $\rho_{t}^{h}(\theta,x,y)$; if
additional contributions to this kernel were considered, with: $\ \left\vert
k^{2}-\lambda_{0}\right\vert \geq2\frac{h}{d_{0}}$, they would generate
contributions to $A_{\theta_{0}}$ allowing exponentially small bounds.

The proof of Theorem~\ref{Th_1} is developed in the Sections \ref{Sec_Krein}
and \ref{Sec_adiabatic}, and summarized at the end of Section~\ref{Sec_proof}.

\section{Spectral properties of the unperturbed Hamiltonian}

The spectral profile of $H_{\theta_{0},\alpha(t)}^{h}(\theta)$, as well as the
dynamics related to, are strictly connected with the Green's kernel and the
generalized eigenfunctions of the corresponding unperturbed operator,
$H_{\theta_{0},0}^{h}(\theta)$. According to the assumption (h1), in what
follows we will focus our attention on the particular case: $\theta=\theta
_{0}=i\tau$\,,%
\begin{equation}
H_{\theta_{0},0}^{h}(\theta_{0})=-h^{2}e^{-2\theta_{0}\,1_{\mathbb{R}%
\backslash(a,b)}}\Delta_{2\theta_{0}}+1_{(a,b)}V_{0}\,, \label{H_unpert}%
\end{equation}
which has been considered, in a more general setting, in \cite{FMN2}. By
making use of the results of Proposition 3.5 in \cite{FMN2} one has:
$\sigma_{ess}(H_{\theta_{0},0}^{h}(\theta_{0}))=e^{-2\theta_{0}}\mathbb{R}%
_{+}$, while a direct computation shows that the point spectrum of
$H_{\theta_{0},0}^{h}(\theta_{0})$ is formed by the solutions to the algebraic
problem%
\begin{equation}
e^{i\frac{\sqrt{z-V_{0}}}{h}(b-a)}=\frac{\sqrt{z-V_{0}}+\sqrt{z}e^{-\theta
_{0}}}{\sqrt{z-V_{0}}-\sqrt{z}e^{-\theta_{0}}},\qquad\arg z\in\left(
-\frac{3}{2}\pi,\frac{\pi}{2}\right)  \,, \label{eigenvalues}%
\end{equation}
fulfilling the condition: $\operatorname{Im}e^{\theta_{0}}\sqrt{z}>0$. These
are explicitly developed in power of $h$ as follows%
\begin{equation}
z_{n}^{h}(\theta_{0})=V_{0}+h^{2}\left(  \frac{n\pi}{b-a}\right)  ^{2}%
-4i\frac{h^{3}}{\left(  b-a\right)  ^{3}}\left(  n\pi\right)  ^{2}%
+h^{4}r_{\theta_{0}}\left(  n\right)  ,\qquad n=0,1,2,...
\label{eigenvalues_1}%
\end{equation}
The remainder is uniformly bounded w.r.t. $\theta_{0}$, for $\left\vert
\theta_{0}\right\vert $ small, while: $\left\vert r_{\theta_{0}}\left(
n\right)  \right\vert =\mathcal{O}\left(  n^{2}\right)  $. In particular, for
$n=0$, the first spectral point coincide with $z_{0}^{h}(\theta_{0})=V_{0}$.

Let $\lambda_{0}\in\left(  \delta,V_{0}-\delta\right)  $ for some $\delta>0$,
and consider the neighbourhood $\mathcal{\tilde{G}}_{h}(\lambda_{0}%
)\subset\subset\left(  0,V_{0}\right)  $%
\begin{equation}
\mathcal{\tilde{G}}_{h}(\lambda_{0})=\left\{  z\in\mathbb{C},\ \left\vert
z-\lambda_{0}\right\vert <Ch,\ \left\vert \arg z\right\vert <h^{N_{0}%
}\right\}  \,, \label{G_h}%
\end{equation}
defined with: $\delta-Ch>0$. For $\theta_{0}=ih^{N_{0}}$ and $z\in
\mathcal{\tilde{G}}_{h}(\lambda_{0})$\,, the result of Proposition 6.5 in
\cite{FMN2} yields the following estimates%
\begin{equation}
\left\Vert \left(  H_{\theta_{0},0}^{h}(\theta_{0})-z\right)  ^{-1}\right\Vert
_{\mathcal{L}\left(  L^{2}(\mathbb{R}),H^{1}(\mathbb{R}\backslash\left\{
a,b\right\}  )\right)  }\leq\frac{C_{a,b,\delta}}{h^{N_{0}+2}}\,,
\label{res_est1}%
\end{equation}%
\begin{equation}
\left\Vert \left(  H_{\theta_{0},0}^{h}(\theta_{0})-z\right)  ^{-1}%
\psi\right\Vert _{\mathcal{L}\left(  H^{-1}(a,b),H^{1}(\mathbb{R}%
\backslash\left\{  a,b\right\}  )\right)  }\leq\frac{C_{a,b,\delta,\psi}%
}{h^{N_{0}+3}}\,, \label{res_est_2}%
\end{equation}
holding for any $\psi\in C_{0}^{\infty}(a,b)$, with constants depending on the data.

The generalized eigenfunctions problem for the incoming waves, in the non
deformed setting $H_{\theta_{0},0}^{h}=-h^{2}\Delta_{\theta_{0}}%
+1_{(a,b)}V_{0}$, writes as%
\begin{equation}
\left(  H_{\theta_{0},0}^{h}-k^{2}\right)  \tilde{\psi}_{-}(k,\cdot)=0,\qquad
k>0\,. \label{gen_eigenf}%
\end{equation}
The exterior part of the solution is%
\begin{equation}
\tilde{\psi}_{-}(k,\cdot)=\left\{
\begin{array}
[c]{l}%
\medskip e^{i\frac{k}{h}x}+R(k)e^{-i\frac{k}{h}x},\qquad\ x<a\\
T(k)e^{i\frac{k}{h}x},\qquad\qquad\qquad x>b
\end{array}
\right.  , \label{gen_eigenf_ext}%
\end{equation}
while, according to the boundary conditions in $D(\Delta_{\theta_{0}})$, the
interior problem is%
\begin{equation}
\left\{
\begin{array}
[c]{l}%
\medskip\left(  -h^{2}\partial_{x}^{2}+V_{0}-k^{2}\right)  \tilde{\psi}%
_{-}(k,\cdot)=0\,,\qquad\text{in }(a,b)\,,\\
\medskip\left(  h\partial_{x}+ike^{-\theta_{0}}\right)  \tilde{\psi}%
_{-}(k,a^{+})=2ike^{i\frac{k}{h}a}e^{-\frac{3}{2}\theta_{0}}\,,\\
\left(  h\partial_{x}-ike^{-\theta_{0}}\right)  \tilde{\psi}_{-}(k,b^{-})=0\,,
\end{array}
\right.
\end{equation}
It follows from a direct computation that%
\begin{equation}
1_{(a,b)}\tilde{\psi}_{-}(k,\cdot)=-\frac{2\sin\gamma_{k^{2}}e^{i\frac{k}{h}%
a}e^{-\frac{\theta_{0}}{2}}}{\sin\left(  \frac{\Lambda_{k^{2}}}{h}%
(b-a)+2\gamma_{k^{2}}\right)  }\cos\left(  \frac{\Lambda_{k^{2}}}%
{h}(x-b)-\gamma_{k^{2}}\right)  \,, \label{gen_eigenf_in}%
\end{equation}
with%
\begin{equation}
\Lambda_{z}=\left(  z-V_{0}\right)  ^{\frac{1}{2}}\,;\qquad e^{2i\gamma_{z}%
}=\frac{\Lambda_{z}-z^{\frac{1}{2}}e^{-\theta_{0}}}{\Lambda_{z}+z^{\frac{1}%
{2}}e^{-\theta_{0}}}\,. \label{small_gamma}%
\end{equation}
The generalized eigenstates of $H_{\theta_{0},0}^{h}(\theta_{0})$ are obtained
by transformation through the deformation map $U_{\theta_{0}}$; in particular,
the interior part of these functions is not affected by the deformation and
one has: $1_{(a,b)}U_{\theta_{0}}\tilde{\psi}_{-}(k,\cdot)=1_{\left(
a,b\right)  }\tilde{\psi}_{-}(k,\cdot)$.

\subsection{The Green's functions of $H_{\theta_{0},0}^{h}(\theta_{0})$}

Assume $z$ to be close to some limit energy $\lambda_{0}$ in the interval
$\left(  0,V_{0}\right)  $: $z\in\mathcal{\tilde{G}}_{h}(\lambda_{0})$. In
this set, we use the square root's branch cut fixed along the positive
imaginary axis (corresponding to: $\arg z\in\left(  -\frac{3}{2}\pi,\frac{\pi
}{2}\right)  $). The integral kernel of $\left(  H_{\theta_{0},0}^{h}%
(\theta_{0})-z\right)  ^{-1}$ is defined by%
\begin{equation}
\left(  H_{\theta_{0},0}^{h}(\theta_{0})-z\right)  G^{z}(\cdot,c)=\delta
_{c}\,. \label{Green_1}%
\end{equation}
Focusing our attention on the case $c\in(a,b)$, $G^{z}(\cdot,c)$ writes as
follows:%
\begin{equation}
\left.  G^{z}(x,c)\medskip\right\vert _{\substack{x\notin\left(  a,b\right)
\\c\in\left(  a,b\right)  }}=\left\{
\begin{array}
[c]{c}%
\medskip u_{+}e^{i\frac{\sqrt{z}e^{\theta_{0}}}{h}(x-b)},\qquad x>b\,,\\
u_{-}e^{-i\frac{\sqrt{z}e^{\theta_{0}}}{h}(x-a)},\qquad x<a\,,
\end{array}
\right.  \label{Green_ext}%
\end{equation}
while the inner problem can be rephrased as%
\begin{equation}
\left\{
\begin{array}
[c]{l}%
\medskip\left(  -h^{2}\partial_{x}^{2}+V_{0}-z\right)  G^{z}(\cdot
,c)=0\,,\qquad\text{for }x\in(a,b)\backslash\left\{  c\right\}  \,,\\
\medskip G^{z}(\cdot,c)\in H^{1}(a,b);\quad h^{2}\left(  \partial_{x}%
G^{z}(c^{+},c)-\partial_{x}G^{z}(c^{-},c)\right)  =-1\,,\\
\medskip\left(  h\partial_{x}+i\sqrt{z}e^{-\theta_{0}}\right)  G^{z}%
(a^{+},c)=0\,,\\
\left(  h\partial_{x}-i\sqrt{z}e^{-\theta_{0}}\right)  G^{z}(b^{-},c)=0\,,
\end{array}
\right.  \label{Green_in.eq}%
\end{equation}
Whit the notation adopted in (\ref{small_gamma}), the solution is%
\begin{equation}
G^{z}(x,c)\medskip=-\frac{1}{h\Lambda_{z}}\frac{1}{\tan\left(  \frac
{\Lambda_{z}}{h}(c-a)+\gamma_{z}\right)  -\tan(\frac{\Lambda_{z}}%
{h}(c-b)-\gamma_{z})}\cdot\left\{
\begin{array}
[c]{c}%
\medskip\frac{\cos\left(  \frac{\Lambda_{z}}{h}(x-a)+\gamma_{z}\right)  }%
{\cos\left(  \frac{\Lambda_{z}}{h}(c-a)+\gamma_{z}\right)  },\qquad
x\in(a,c)\,,\\
\frac{\cos\left(  \frac{\Lambda_{z}}{h}(x-b)-\gamma_{z}\right)  }{\cos\left(
\frac{\Lambda_{z}}{h}(c-b)-\gamma_{z}\right)  },\qquad x\in(c,b)\,.
\end{array}
\right.  \label{Green_in}%
\end{equation}
It follows from the definition of $\mathcal{\tilde{G}}_{h}(\lambda_{0})$ and
our choice of the branch cut, that: $\operatorname{Im}\sqrt{z}e^{\theta_{0}%
}>0$. Thus (\ref{Green_ext}) and (\ref{Green_in}), properly defines $L^{2}%
$-functions. When $\operatorname{Re}z\in\left(  0,V_{0}\right)  $ -- as is the
case for $z\in\mathcal{\tilde{G}}_{h}(\lambda_{0})$ -- the point $z-V_{0}$ has
a negative real part and, according to the definition of the square root, one
has: $\operatorname{Im}\Lambda_{z}<0$. Therefore, the terms of the type:
$e^{i\frac{\Lambda_{z}}{h}\,d}$ appearing in (\ref{Green_in}) are
exponentially increasing or decreasing functions of $\frac{1}{h}$ depending on
the sign of $d$. The asymptotic behaviour of the value $G^{z}(c,c)$ as
$h\rightarrow0$ will be considered by using the formula: $l=(b-a)$,
$p_{\theta_{0}}(z)=e^{-2i\gamma_{z}}$%
\begin{equation}
G^{z}(c,c)=-\frac{i}{2h\Lambda_{z}}\frac{1}{1-e^{-2i\frac{\Lambda_{z}}{h}%
l}p_{\theta_{0}}^{2}(z)}\left[  1+e^{-2i\frac{\Lambda_{z}}{h}(c-a)}%
p_{\theta_{0}}(z)+e^{-2i\frac{\Lambda_{z}}{h}(b-c)}p_{\theta_{0}%
}(z)+e^{-2i\frac{\Lambda_{z}}{h}l}p_{\theta_{0}}^{2}(z)\right]
\label{Green_exp}%
\end{equation}
which is a rewriting of (\ref{Green_in}).

\begin{remark}
\label{Rem_1}Although $G^{z}$ is properly defined for $z\in\mathbb{C}%
\backslash\mathbb{R}_{+}e^{-2\theta_{0}}\cup\left\{  z_{n}^{h}(\theta
_{0})\right\}  $, and in particular for $z\in\mathcal{\tilde{G}}_{h}%
(\lambda_{0})$, the relation (\ref{Green_exp}) makes sense in $\mathbb{C}%
\backslash\left\{  z_{n}^{h}(\theta_{0})\right\}  $. Thus, considering the
small-$h$ expansions of $G^{z}(c,c)$, a larger neighbourhood of $\lambda_{0}$
can be used, such that: $\operatorname{Im}\Lambda_{z}<0$.
\end{remark}

Next we give accurate upper and lower bounds and exponential estimates, for
$G^{z}(\cdot,c)$\textbf{, }$c\in\left(  a,b\right)  $. Using (\ref{Green_ext}%
)-(\ref{Green_in}), allows to show that this function is exponentially
decaying outside a small neighbourhood of $x=c$\,, where all its mass
concentrates as $h\rightarrow0$.

\begin{lemma}
\label{Lemma_1}Let $z\in\mathcal{\tilde{G}}_{h}(\lambda_{0})$, $\theta
_{0}=ih^{N_{0}}$ and $h\in\left(  0,h_{0}\right)  $ with $h_{0}$ small. For
$c\in\left(  a,b\right)  $, the following estimates holds%
\begin{equation}
\frac{c_{0}}{h}\leq\left\Vert G^{z}(\cdot,c)\right\Vert _{L^{2}(a,b)}^{2}%
\leq\frac{c_{1}}{h}\,, \label{Green_uplow}%
\end{equation}%
\begin{equation}
\sup_{\left[  a,b\right]  }\left\vert e^{\frac{^{\varphi}}{h}}G^{z}%
(\cdot,c)\right\vert +\left\Vert e^{\frac{^{\varphi}}{h}}h\partial_{x}%
G^{z}(\cdot,c)\right\Vert _{L^{2}(a,b)}+\left\Vert e^{\frac{^{\varphi}}{h}%
}G^{z}(\cdot,c)\right\Vert _{L^{2}(a,b)}\leq\frac{C_{a,b}}{h}\,,
\label{Green_exp_est}%
\end{equation}%
\begin{equation}
\left\Vert G^{z}(\cdot,c)\right\Vert _{L^{2}(\mathbb{R}\backslash(a,b))}\leq
C_{a,b}h^{-\frac{N_{0}+1}{2}}e^{-\frac{\beta(\lambda_{0})}{h}}\,,
\label{Green_est_ext2}%
\end{equation}%
\begin{equation}
\left\Vert G^{z_{1}}-G^{z_{2}}\right\Vert _{L^{2}(\mathbb{R})}\leq
\frac{C_{a,b,\delta}}{h^{N_{0}+2}}\,\left\vert z_{1}-z_{2}\right\vert
\,;\qquad\left\Vert \partial_{z}G^{z}(\cdot,c)\right\Vert _{L^{2}(\mathbb{R}%
)}\leq\frac{C_{a,b,c,\delta}}{h^{N_{1}}}\,, \label{Green_der_est}%
\end{equation}
with: $\varphi=(V_{0}-\lambda_{0})^{\frac{1}{2}}\left\vert \cdot-c\right\vert
$, $\beta(\lambda_{0})=(V_{0}-\lambda_{0})^{\frac{1}{2}}\,d(c,\left\{
a,b\right\}  )$, $N_{1}=2N_{0}+5$ and constants depending on the data.
\end{lemma}

\begin{preuve}
To simplify the notation, we use $G^{z}$ instead of $G^{z}(\cdot,c)$. Let
start considering an upper bound of $\left\Vert G^{z}\right\Vert _{L^{2}%
(a,b)}$ in the relevant energy range: $\operatorname{Re}z\in\left(
0,V_{0}\right)  $. Owing to (\ref{Green_in}), we have%
\[
\left.  G^{z}\medskip\right\vert _{x\in(a,b)\backslash\left\{  c\right\}
}=\frac{1}{ih\Lambda_{z}}e^{-i\frac{\Lambda_{z}}{h}\left\vert x-c\right\vert
}\left[  1+\mathcal{O}\left(  e^{-2i\frac{\Lambda_{z}}{h}d(x,\left\{
a,b\right\}  )}\right)  \right]  \,.
\]
Since $\operatorname{Im}\Lambda_{z}<0$ when $\operatorname{Re}z\in\left(
0,V_{0}\right)  $, the quantities: $e^{-i\frac{\Lambda_{z}}{h}\left\vert
x-c\right\vert }$, $e^{-2i\frac{\Lambda_{z}}{h}d(x,\left\{  a,b\right\}  )}$
are exponentially small as $h\rightarrow0$, and an explicit computation yields%
\begin{equation}
\left\Vert G^{z}(\cdot,c)\right\Vert _{L^{2}(a,b)}^{2}\leq\frac{C_{1}%
}{h\left\vert \operatorname{Im}\Lambda_{z}\right\vert }\leq\frac{\tilde{C}%
_{1}}{h}\,. \label{Green_upb}%
\end{equation}
For the lower bound, we refer to (\ref{Green_in}), with $x\in\left(
a,c\right)  $, to get%
\[
\left.  G^{z}(x)\right\vert _{x\in(a,c)}\geq\frac{C_{0}}{h}\left\vert
\frac{\cos\left(  \frac{\Lambda_{z}}{h}(x-a)+\gamma_{z}\right)  }{\cos\left(
\frac{\Lambda_{z}}{h}(c-a)+\gamma_{z}\right)  }\right\vert \geq\frac{\tilde
{C}_{0}}{h}\left\vert e^{-i\frac{\Lambda_{z}}{h}(c-x)}\right\vert \,,
\]
for $\delta$ and $h$ small. It follows%
\begin{equation}
\medskip\left\Vert G^{z}\right\Vert _{L^{2}(a,b)}^{2}\geq\left\Vert
G^{z}\right\Vert _{L^{2}(c-\delta,c)}^{2}\geq\frac{\tilde{C}_{0}}{h^{2}}%
\int_{c-\delta}^{c}\left\vert e^{-i\frac{\Lambda_{z}}{h}(c-x)}\right\vert
^{2}dx\geq\frac{\tilde{C}_{0}}{4h\left\vert \operatorname{Im}\Lambda
_{z}\right\vert }\,. \label{Green_lowb}%
\end{equation}
Exponential estimates are usually obtained from Agmon identities using as
exponential weight the distance from the classical region of motion (we refer
to \cite{Hel}). However, in this particular case, the explicit formula
(\ref{Green_in}) shows that the leading factors in $G^{z}$ and $h\partial
_{x}G^{z}$ are controlled by: $e^{-i\frac{\Lambda_{z}}{h}\left\vert
x-c\right\vert }\lesssim e^{-\frac{(V_{0}-\lambda_{0})^{\frac{1}{2}}}%
{h}\left\vert x-c\right\vert }$; this gives (\ref{Green_exp_est}). In the
exterior domain, a direct computation yields%
\begin{equation}
\left\Vert G^{z}\right\Vert _{L^{2}(\mathbb{R}\backslash(a,b))}^{2}=\frac
{h}{2\operatorname{Im}\left(  ze^{2\theta_{0}}\right)  ^{\frac{1}{2}}}\left(
\left\vert G^{z}(a^{+})\right\vert ^{2}+\left\vert G^{z}(b^{-})\right\vert
^{2}\right)  \,.
\end{equation}
For $z\in\mathcal{\tilde{G}}_{h}(\lambda_{0})$, $\operatorname{Im}\left(
ze^{2\theta_{0}}\right)  ^{\frac{1}{2}}\sim\mathcal{O}(h^{N_{0}})$, and we get%
\begin{equation}
\left\Vert G^{z}\right\Vert _{L^{2}(\mathbb{R}\backslash(a,b))}^{2}\leq
\frac{1}{2h^{N_{0}-1}}\left(  \left\vert G^{z}(a^{+})\right\vert
^{2}+\left\vert G^{z}(b^{-})\right\vert ^{2}\right)  \,.
\label{Green_est_ext1}%
\end{equation}
According to (\ref{Green_exp_est}), the boundary values of $G^{z}$ are
estimated by $\frac{e^{-\frac{\beta\left(  \lambda_{0}\right)  }{h}}}{h}$ and
the inequality (\ref{Green_est_ext2}) follows.

From the definition (\ref{Green_ext})-(\ref{Green_in}), $z\rightarrow G^{z}$
is an $L^{2}$-valued holomorphic map in $\mathcal{\tilde{G}}_{h}(\lambda_{0}%
)$. For $z_{1},z_{2}\in\mathcal{\tilde{G}}_{h}(\lambda_{0})$, we have%
\begin{align}
\medskip G^{z_{1}}-G^{z_{2}}  &  =\left[  \left(  H_{\theta_{0},0}^{h}%
(\theta_{0})-z_{1}\right)  ^{-1}-\left(  H_{\theta_{0},0}^{h}(\theta
_{0})-z_{2}\right)  ^{-1}\right]  \delta_{c}\nonumber\\
&  =(z_{1}-z_{2})\left(  H_{\theta_{0},0}^{h}(\theta_{0})-z_{1}\right)
^{-1}\left(  H_{\theta_{0},0}^{h}(\theta_{0})-z_{2}\right)  ^{-1}\delta
_{c}=(z_{1}-z_{2})\,\left(  H_{\theta_{0},0}^{h}(\theta_{0})-z_{1}\right)
^{-1}G^{z_{2}}\,. \label{Green_diff}%
\end{align}
Using (\ref{res_est1}), the first of (\ref{Green_der_est}) follows. The
derivative $\partial_{z}G^{z}$, is expressed by%
\begin{equation}
\partial_{z}G^{z}=\left(  H_{\theta_{0},0}^{h}(\theta_{0})-z\right)
^{-1}G^{z}\,, \label{Green_der}%
\end{equation}
which implies: $\partial_{z}G^{z}=\left(  H_{\theta_{0},0}^{h}(\theta
_{0})-z\right)  ^{-2}\delta_{c}$. Then, using (\ref{res_est1}%
)-(\ref{res_est_2}) completes the proof of (\ref{Green_der_est}).

\hfill
\end{preuve}

\section{\label{Sec_Krein}A Krein's resolvent formula and spectral expansions
for small $h$}

Let consider the spectral problem for the deformed Hamiltonian with
$\theta=\theta_{0}=ih^{N_{0}}$,%
\begin{equation}
H_{\theta_{0},\alpha}^{h}(\theta_{0})=-h^{2}e^{-2\theta_{0}\,1_{\mathbb{R}%
\backslash(a,b)}}\Delta_{2\theta_{0}}+1_{(a,b)}V_{0}+h\alpha\,\delta_{c},\quad
c\in\left(  a,b\right)  \,. \label{H_pert}%
\end{equation}
The related resolvent operator can be expressed as a finite rank perturbation
of $\left(  H_{\theta_{0},0}^{h}(\theta_{0})-z\right)  ^{-1}$%
\begin{equation}
\left(  H_{\theta_{0},\alpha}^{h}(\theta_{0})-z\right)  ^{-1}=\left(
H_{\theta_{0},0}^{h}(\theta_{0})-z\right)  ^{-1}-\frac{h\alpha\,\left\langle
\bar{G}^{z}(\cdot,c),\cdot\right\rangle _{L^{2}(R)}}{1+h\alpha\,G^{z}%
(c,c)}\,G^{z}(\cdot,c)\,. \label{Krein}%
\end{equation}
This will provide an accurate description of the resonant energy as
$h\rightarrow0$.

\begin{proposition}
\label{Prop_1}Let $h\in\left(  0,h_{0}\right)  $, with $h_{0}$ small,
$\theta_{0}=ih^{N_{0}}$ and $\alpha\in\left(  -2V_{0}^{\frac{1}{2}},0\right)
$. The spectrum of $H_{\theta_{0},\alpha}^{h}(\theta_{0})$ is characterized by
the conditions:\medskip\newline$i)$ The essential spectrum is: $\sigma
_{ess}(H_{\theta_{0},\alpha}^{h}(\theta_{0}))=e^{-2\theta_{0}}\mathbb{R}_{+}%
$.\medskip\newline$ii)$ There exists a unique nondegenerate spectral point of
$H_{\theta_{0},\alpha}^{h}(\theta_{0})$ in $\left\{  \operatorname{Re}%
z\in\left(  0,V_{0}\right)  ,\ \arg z\in\left(  -2\theta_{0},0\right)
\right\}  $, admitting the small-$h$ expansion%
\begin{equation}
E_{res}^{h}=V_{0}-\frac{\alpha^{2}}{4}-\frac{\alpha^{2}}{2}p_{0}%
(E^{0})e^{-\frac{\left\vert \alpha\right\vert }{h}d(c,\left\{  a,b\right\}
)}+\mathcal{O}\left(  \theta_{0}\,e^{-\frac{\left\vert \alpha\right\vert }%
{h}d(c,\left\{  a,b\right\}  )}\right)  +o(e^{-\frac{\left\vert \alpha
\right\vert }{h}d(c,\left\{  a,b\right\}  )})\,, \label{eigenvalue_exp}%
\end{equation}
with: $p_{0}(E^{0})=\frac{1}{V_{0}}\left[  i\left\vert \alpha\right\vert
\left(  V_{0}-\frac{\alpha^{2}}{4}\right)  ^{\frac{1}{2}}-\left(  V_{0}%
-\frac{\alpha^{2}}{2}\right)  \right]  $. The corresponding eigenvector is
given by the Green's function: $G^{E_{res}^{h}}(\cdot,c)$.\medskip
\newline$iii)$ Both $E_{res}^{h}$ and $G^{E_{res}^{h}}(\cdot,c)$ are
holomorphic w.r.t. $\alpha$.
\end{proposition}

\begin{preuve}
$i)$ The first statement is a consequence of Corollary 3.4 in \cite{FMN2}
(holding for generic $\mathcal{M}_{b}$-perturbation of $H_{\theta_{0},0}%
^{h}(\theta_{0})$ supported in $\left(  a,b\right)  $).

$ii)$ According to Proposition 5.5 in \cite{FMN2} (partly relying on
Helffer-Sj\"{o}strand techniques in \cite{HeSj1}), the points in $\left\{
\operatorname{Re}z\in\left(  0,V_{0}\right)  ,\ \arg z\in\left(  -2\theta
_{0},0\right)  \right\}  \cap\sigma(H_{\theta_{0},0}^{h}(\theta_{0}))$ are
localized around the eigenvalues of the Dirichlet Hamiltonian: $H_{D}%
^{h}=-\Delta_{\left(  a,b\right)  }^{D}+1_{\left(  a,b\right)  }V_{0}%
+\alpha\,\delta_{c}$, with a one to one correspondence and an exponentially
small bound. Using the dilation: $x\rightarrow\frac{x-c}{h}$, we can refer to
the spectral problem for $H_{d}^{h}=-\Delta_{\left(  \frac{\tilde{a}}{h}%
,\frac{\tilde{b}}{h}\right)  }^{D}+1_{\left(  \frac{\tilde{a}}{h},\frac
{\tilde{b}}{h}\right)  }V_{0}+\alpha\,\delta_{c}$. When $h\rightarrow0$, the
spectral subset $\left(  0,V_{0}\right)  \cap\sigma(H_{d}^{h})$ converges to
$\left(  0,V_{0}\right)  \cap\sigma(H_{d}^{0})$, with: $H^{0}=-\Delta
+V_{0}+\alpha\,\delta_{c}$, preserving the dimension of the respective
subspaces (the proof of this point is based on standard convergence estimates
in semiclassical analysis; a guide line for it can be recovered from the
strategy used in \cite{FMN1} Lemma 4.5). The point spectrum of $H^{0}$ is
explicitly computable: for $\alpha\in\left(  -2V_{0}^{\frac{1}{2}},0\right)
$, it is composed of a unique point of multiplicity 1 and equal to
$\lambda=V_{0}-\frac{\left\vert \alpha\right\vert ^{2}}{4}$. Therefore, there
exists a unique non-degenerate eigenvalue, $E_{res}^{h}$, of $H_{\theta
_{0},\alpha}^{h}(\theta_{0})$ in the prescribed region, converging to
$\lambda$ as $h\rightarrow0$\,. The result of Proposition 5.5 in \cite{FMN2},
writes in this case as: $E_{res}^{h}=V_{0}-\frac{\alpha^{2}}{4}+\mathcal{O}%
(h^{-3}e^{-\frac{\left\vert \alpha\right\vert }{h}d(c,\left\{  a,b\right\}
)})$.

A more refined asymptotic expression is obtained by using the explicit
resolvent's formula. Since the poles of $\left(  H_{\theta_{0},0}^{h}%
(\theta_{0})-z\right)  ^{-1}$ are confined in $\left\{  \operatorname{Re}z\geq
V_{0}\right\}  $ (we refer to (\ref{eigenvalues_1})), the relation
(\ref{Krein}) leads to an equation for $E_{res}^{h}$%
\begin{equation}
1+h\alpha\,G^{E}(c,c)=0 \label{eigenvalue_eq1}%
\end{equation}
In the strip $\operatorname{Re}E\in\left(  0,V_{0}\right)  $, where:
$\operatorname{Im}\left(  E-V_{0}\right)  ^{\frac{1}{2}}<0$ due to the
determination $\arg z\in\left(  -\frac{3}{2}\pi,\frac{\pi}{2}\right)  $, the
above equation is rephrased as%
\begin{gather}
2\left(  E-V_{0}\right)  ^{\frac{1}{2}}-\frac{i\alpha}{1-e^{-2i\frac{\left(
E-V_{0}\right)  ^{\frac{1}{2}}}{h}l}p_{\theta_{0}}^{2}(E)}\left[
1+e^{-2i\frac{\left(  E-V_{0}\right)  ^{\frac{1}{2}}}{h}(c-a)}p_{\theta_{0}%
}(E)\right.  \qquad\qquad\qquad\qquad\qquad\qquad\qquad\qquad\qquad
\qquad\nonumber\\
\qquad\qquad\qquad\qquad\qquad\left.  +e^{-2i\frac{\left(  E-V_{0}\right)
^{\frac{1}{2}}}{h}(b-c)}p_{\theta_{0}}(E)+e^{-2i\frac{\left(  E-V_{0}\right)
^{\frac{1}{2}}}{h}l}p_{\theta_{0}}^{2}(E)\right]  =0\,, \label{eigenvalue_eq2}%
\end{gather}
according to (\ref{Green_exp}). Let $E=V_{0}-\frac{\alpha^{2}}{4}+\delta
E\,e^{-\frac{\left\vert \alpha\right\vert }{h}d(c,\left\{  a,b\right\}
)}+o(e^{-\frac{\left\vert \alpha\right\vert }{h}d(c,\left\{  a,b\right\}  )}%
)$; an approximation of the first order in $e^{-\frac{\left\vert
\alpha\right\vert }{h}d(c,\left\{  a,b\right\}  )}$ of (\ref{eigenvalue_eq2})
yields%
\begin{equation}
\delta E\,=-\frac{\alpha^{2}}{2}p_{\theta_{0}}(E^{0})\,,
\label{eigenvalue_exp_1}%
\end{equation}
where $p_{\theta_{0}}(E)$ is holomorphic w.r.t. both the variables provided
that $E\sim V_{0}-\frac{\alpha^{2}}{4}$ and $\left\vert \theta_{0}\right\vert
<<1$. The value $p_{\theta_{0}}(E^{0})$ is approximated by: $p_{\theta_{0}%
}(E^{0})=\gamma_{0}(E^{0})+\mathcal{O}(\theta_{0})$,%
\[
p_{0}(E^{0})=\frac{1}{V_{0}}\left[  i\left\vert \alpha\right\vert \left(
V_{0}-\frac{\alpha^{2}}{4}\right)  ^{\frac{1}{2}}-\left(  V_{0}-\frac
{\alpha^{2}}{2}\right)  \right]  \,,
\]
this leads to the expansion (\ref{eigenvalue_exp}). Finally, the relation
(\ref{eigenvalue_eq1}), allows to verify that:\newline$\left(  H_{\theta
_{0},\alpha}^{h}(\theta_{0})-E_{res}^{h}\right)  G^{E_{res}^{h}}(\cdot,c)=0$,

$iii)$ Consider the quadratic form associated with the operator $H_{\theta
_{0},\alpha}^{h}(\theta_{0})$. This is an accretive form (due to the choice
$\theta=\theta_{0}$) and its domain, $Q(H_{\theta_{0},0}^{h}(\theta
_{0}))=H^{2}\left(  \mathbb{R}\backslash\left\{  a,c,b\right\}  \right)  \cup
H^{1}\left(  \mathbb{R}\backslash\left\{  a,b\right\}  \right)  $, is
independent of $h$ and $\alpha$. For any $u\in Q(H_{\theta_{0},0}^{h}%
(\theta_{0})$, its action%
\[
\left\langle u,\,H_{\theta_{0},\alpha}^{h}(\theta_{0})u\right\rangle
_{L^{2}(\mathbb{R})}=h^{2}\sin2\theta_{0}\int_{\mathbb{R}\backslash\left(
a,b\right)  }\left\vert u^{\prime}\right\vert ^{2}+V_{0}\int_{\mathbb{R}%
}\left\vert u\right\vert ^{2}+h\alpha\left\vert u(c)\right\vert ^{2}%
\]
defines an holomorphic function of $\alpha$. Thus, $H_{\theta_{0},\alpha}%
^{h}(\theta_{0})$ is an analytic family type $B$ w.r.t. $\alpha$ and the
Kato-Rellich Theorem applies to the non-degenerate discrete eigenvalue
$E_{res}^{h}$.
\end{preuve}

\begin{remark}
The solution $E_{res}^{h}$ is a singularity of the resolvent embedded in the
second Riemann sheet and corresponds to the shape-resonance produced by the
attractive part of the interaction. It defines an eigenvalue of the deformed
operator $H_{\theta_{0},\alpha}^{h}(\theta_{0})$ provided that $\left\vert
\arg E_{res}^{h}\right\vert $ is lower than the deformation angle, given by
$h^{N_{0}}$ in our assumption. Since $\operatorname{Im}E_{res}^{h}$ is
exponentially small w.r.t. $h$ this condition definitively holds as
$h\rightarrow0$.
\end{remark}

Next, we consider the expansions of relevant quantities involved in the
computation of $A_{\theta_{o}}(t)$ for energies close to the resonance. In
what follows, we assume the results of Proposition \ref{Prop_1} to hold, and
take $\left\vert E-E_{res}^{h}\right\vert <\frac{h}{d}$, $d$ being a small
constant which fixes a complex neighbourhood of $E_{res}^{0}$ of size $h$. In
such a domain, the relation (\ref{Green_exp}) can be used to write expansions
of $G^{E}(c,c)$ as $h\rightarrow0$ (see Remark \ref{Rem_1}). Recalling that:
$1+h\alpha G^{E_{res}^{h}}(c,c)=0$, the function $\left(  1+h\alpha
G^{E}(c,c)\right)  ^{-1}$ can be written in the form%
\begin{equation}
\left(  1+h\alpha G^{E}(c,c)\right)  ^{-1}=\frac{M(E,E_{res}^{h})}%
{E-E_{res}^{h}}\,, \label{exp_1.1}%
\end{equation}%
\begin{equation}
M(E,E_{res}^{h})=\frac{E-V_{0}+\left(  E_{res}^{h}-V_{0}\right)  ^{\frac{1}%
{2}}\left(  E-V_{0}\right)  ^{\frac{1}{2}}}{1+h\alpha\left[  \left(
E-V_{0}\right)  ^{\frac{1}{2}}+\left(  E_{res}^{h}-V_{0}\right)  ^{\frac{1}%
{2}}\right]  \frac{\left(  E-V_{0}\right)  ^{\frac{1}{2}}G^{E}(c,c)-\left(
E_{res}^{h}-V_{0}\right)  ^{\frac{1}{2}}G^{E_{res}^{h}}(c,c)}{E-E_{res}^{h}}%
}\,, \label{M_E_def}%
\end{equation}
with the branch-cut fixed along $i\mathbb{R}_{+}$. The incremental ratio at
the denominator in (\ref{M_E_def}) is controlled by the derivative of $\left(
E-V_{0}\right)  ^{\frac{1}{2}}G^{E}(c)$ evaluated in a neighbourhood of
$E_{res}^{h}$. According to (\ref{Green_exp}), this writes as%
\begin{gather*}
\left(  E-V_{0}\right)  ^{\frac{1}{2}}G^{E}(c,c)=-\frac{i}{2h}\frac
{1}{1-e^{-2i\frac{\left(  E-V_{0}\right)  ^{\frac{1}{2}}}{h}l}p_{\theta_{0}%
}^{2}(E)}\left[  1+e^{-2i\frac{\left(  E-V_{0}\right)  ^{\frac{1}{2}}}%
{h}(c-a)}p_{\theta_{0}}(E)\right.  \qquad\qquad\qquad\qquad\\
\qquad\qquad\qquad\qquad\qquad\qquad\qquad\qquad\qquad\qquad\left.
+e^{-2i\frac{\left(  E-V_{0}\right)  ^{\frac{1}{2}}}{h}(b-c)}p_{\theta_{0}%
}(E)+e^{-2i\frac{\left(  E-V_{0}\right)  ^{\frac{1}{2}}}{h}l}p_{\theta_{0}%
}^{2}(E)\right]  \,.
\end{gather*}
Using the holomorphicity of $p_{\theta_{0}}(E)$ and $\left(  E-V_{0}\right)
^{\frac{1}{2}}$ in $\left\{  \left\vert E-E_{res}^{h}\right\vert <\frac{h}%
{d}\right\}  $, and the asymptotic characterization (\ref{eigenvalue_exp}), we
get:%
\begin{equation}
\partial_{E}\left(  E-V_{0}\right)  ^{\frac{1}{2}}G^{E}(c,c)=h^{-2}%
e^{-\frac{\left\vert \alpha\right\vert }{h}d(c,\left\{  a,b\right\}
)}\mathcal{R}(E)\,. \label{M_E_exp}%
\end{equation}
This yields the representation%
\begin{equation}
M(E,E_{res}^{h})=\left[  E-V_{0}+\left(  E_{res}^{h}-V_{0}\right)  ^{\frac
{1}{2}}\left(  E-V_{0}\right)  ^{\frac{1}{2}}\right]  +h^{-1}e^{-\frac
{\left\vert \alpha\right\vert }{h}d(c,\left\{  a,b\right\}  )}\mathcal{R}%
(E)\,, \label{M_E}%
\end{equation}
holding for $\left\vert E-E_{res}^{h}\right\vert <\frac{h}{d}$. In a closer
neighbourhood of the resonance, the function $M(E,E_{res}^{h})$ is connected
with scalar products of Green's functions.

\begin{lemma}
\label{Lemma_2}In the assumptions of Proposition \ref{Prop_1}, let
$E\in\mathcal{\tilde{G}}_{h}(E_{res}^{h})$, $c\in\left(  a,b\right)  $,
$S=\left\vert \alpha\right\vert \,d(c,\left\{  a,b\right\}  )$. The relations
\begin{equation}
h\alpha\left\langle G^{E^{\ast}}(\cdot,c),G^{E_{res}^{h}}(\cdot
,c)\right\rangle _{L^{2}(\mathbb{R})}=\frac{1}{M(E,E_{res}^{h})}%
+h^{-N}e^{-\frac{S}{h}}\mathcal{R}_{0}(E,h)\,, \label{Green_prod}%
\end{equation}%
\begin{equation}
h\alpha\left.  \left\langle G^{E^{\ast}}(\cdot,c),\partial_{z}G^{z}%
(\cdot,c)\right\rangle _{L^{2}(\mathbb{R})}\smallskip\right\vert
_{z=E_{res}^{h}}=\left.  \partial_{z}\frac{1}{M(E,z)}\right\vert
_{z=E_{res}^{h}}+h^{-N_{1}}e^{-\frac{S}{h}}\mathcal{R}_{1}(E,h)\,,
\label{Green_prod_diff}%
\end{equation}
hold with: $N,N_{1}$ suitable positive integers, while $\mathcal{R}_{i}(E,h)$
are holomorphic functions of $E$ uniformly bounded w.r.t. $h$.
\end{lemma}

\begin{preuve}
From the relations $h\alpha G^{E_{res}^{h}}(c,c)=-1$ and (\ref{Green_diff}),
it follows%
\begin{equation}
\left(  1+h\alpha G^{E}(c,c)\right)  =h\alpha\left(  G^{E}(c,c)-G^{E_{res}%
^{h}}(c,c)\right)  =h\alpha(E-E_{res}^{h})\,\left(  H_{\theta_{0},0}%
^{h}(\theta_{0})-E\right)  ^{-1}G^{E_{res}^{h}}(c,c)\,,
\end{equation}
which also writes as%
\begin{equation}
\left(  1+h\alpha G^{E}(c,c)\right)  =h\alpha(E-E_{res}^{h})\,\int
_{\mathbb{R}}G^{E}(c,x)G^{E_{res}^{h}}(x,c)\,dx\,. \label{Green_prod_1}%
\end{equation}
For $x\in\left(  a,b\right)  $, $G^{E}(c,x)$ is obtained from (\ref{Green_in})
by interchanging the variables $x$ and $c$; from a direct check on this
formula it follows: $G^{E}(x,c)=G^{E}(c,x)$. For any $x\in\mathbb{R}%
\backslash\left(  a,b\right)  $, the map $c\rightarrow G^{E}(c,x)$ is the
solution of: $\left(  H_{\theta_{0},0}^{h}(\theta_{0})-z\right)  G^{z}%
(\cdot,x)=\delta_{x}$ in $\left(  a,b\right)  $. According to the boundary
conditions in (\ref{Laplacian_mod}), this problem explicitly writes as%
\begin{equation}
\left\{
\begin{array}
[c]{lll}%
\medskip\left(  -h^{2}\partial_{c}^{2}+V_{0}-E\right)  G^{E}(c,x)=0\qquad &
\text{for }c\in\left(  a,b\right)  \text{\thinspace},\ x>b & \text{for }%
c\in\left(  a,b\right)  \,,\ x<a\,,\\
\left(  h\partial_{c}+i\sqrt{E}e^{-\theta_{0}}\right)  G^{E}(a^{+},x)= & 0 &
-\frac{e^{-\theta_{0}}}{h}e^{i\frac{\sqrt{E}e^{\theta_{0}}}{h}\left(
a-x\right)  }\,,\\
\left(  h\partial_{c}-i\sqrt{E}e^{-\theta_{0}}\right)  G^{E}(b^{-},x)= &
\frac{e^{-\theta_{0}}}{h}e^{i\frac{\sqrt{E}e^{\theta_{0}}}{h}\left(
x-b\right)  } & 0\,.
\end{array}
\right.  \label{Green_c_ext1}%
\end{equation}
For $E\in\mathcal{\tilde{G}}_{h}(E_{res}^{h})$, the Lemma 4.3 in \cite{FMN2}
applies, and a point-wise exponential estimate for $1_{\left(  a,b\right)
}G^{E}(\cdot,x)$ holds, depending on $x$:%
\begin{equation}
\sup_{c\in\left[  a,b\right]  }\left\vert e^{\frac{\varphi}{h}}G^{E}%
(c,x)\right\vert \leq\frac{C_{a,b}}{h^{2}}\left(  \left\vert 1_{x>b}%
e^{i\frac{\sqrt{E}e^{\theta_{0}}}{h}\left(  x-b\right)  }\right\vert
+\left\vert 1_{x<a}e^{i\frac{\sqrt{E}e^{\theta_{0}}}{h}\left(  a-x\right)
}\right\vert \right)  \,, \label{Green_c_ext_exp_est}%
\end{equation}
with: $\varphi=\frac{\left\vert \alpha\right\vert }{2}\,d(\cdot,\left\{
a,b\right\}  )$. From $i)$, the integral at the r.h.s. of (\ref{Green_prod_1})
writes as
\begin{align*}
\int_{\mathbb{R}}G^{E}(c,x)G^{E_{res}^{h}}(x,c)\,dx  &  =\int_{a}^{b}%
G^{E}(x,c)G^{E_{res}^{h}}(x,c)\,dx+\int_{\mathbb{R}\backslash\left(
a,b\right)  }G^{E}(c,x)G^{E_{res}^{h}}(x,c)\,dx\\
&  =\int_{\mathbb{R}}G^{E}(x,c)G^{E_{res}^{h}}(x,c)\,dx-\int_{\mathbb{R}%
\backslash\left(  a,b\right)  }\left[  G^{E}(x,c)-G^{E}(c,x)\right]
G^{E_{res}^{h}}(x,c)\,dx\,.
\end{align*}
The exterior contribution, defines an holomorphic function of $E\in
\mathcal{\tilde{G}}_{h}(E_{res}^{h})$. From (\ref{Green_c_ext_exp_est}) the
Cauchy-Schwarz inequality and the estimate (\ref{Green_est_ext2}), applied
with $\lambda_{0}=V_{0}-\frac{\alpha^{2}}{4}$, this is bounded by
$\mathcal{O}\left(  h^{-N}e^{-\frac{S}{h}}\right)  $ for a suitable large $N$,
uniformly w.r.t. $E$. Then (\ref{Green_prod}) is deduced from (\ref{exp_1.1}).
The second relation (\ref{Green_prod_diff}) similarly follows by using
(\ref{Green_der_est}). \hfill
\end{preuve}

\section{\label{Sec_adiabatic}Adiabatic evolution of $A_{\theta_{0}}(t)$.}

We consider the asymptotic behaviour of the dynamical system (\ref{A_def}%
)-(\ref{A_def2}) as $h\rightarrow0$ goes to zero. The assumptions (h1)-(h4)
fix the physical data of the problem, including: 1) the quantum observable
$\chi$, corresponding to the charge accumulating around the interaction point
$c$\thinspace; 2) the time-profile of the interaction, $\alpha(t)$, which
determines the resonant energy level at time $t$\thinspace;3) the energy
partition function $g$, defining an out-of-equilibrium initial state; 4) the
long time scale of the problem, corresponding to the inverse of the adiabatic
parameter $\varepsilon$ defined in (\ref{eps}).

In particular, the constraint $\alpha_{t}\in\left(  -2V_{0}^{\frac{1}{2}%
},0\right)  $ implies that the attractive part of the interaction generates,
for each $t$, a single resonance whose small-$h$ expansion is given in
(\ref{eigenvalue_exp}). With the notation introduced in Section
\ref{Sec_model}, this corresponds to: $E(t)=E_{R}(t)-i\Gamma_{t}$%
\begin{equation}
E_{R}(t)=V_{0}-\frac{\alpha_{t}^{2}}{4}+\mathcal{O}\left(  e^{-\frac
{\left\vert \alpha_{t}\right\vert }{h}\,d(c,\left\{  a,b\right\}  )}\right)
\label{Res_real}%
\end{equation}%
\begin{equation}
\Gamma_{t}=\mathcal{O}\left(  e^{-\frac{\left\vert \alpha_{t}\right\vert }%
{h}\,d(c,\left\{  a,b\right\}  )}\right)  \label{Res_imm}%
\end{equation}
Due to (\ref{alpha_var}), exponentially small terms $\mathcal{O}\left(
e^{-\frac{\left\vert \alpha_{t}\right\vert }{h}\,d(c,\left\{  a,b\right\}
)}\right)  $ can be replaced by $\mathcal{O}\left(  e^{-\frac{\left\vert
\alpha_{0}\right\vert }{h}\,d(c,\left\{  a,b\right\}  )}\right)  $; this leads
to%
\begin{equation}
\Gamma_{t}=\mathcal{O}\left(  e^{-\frac{\left\vert \alpha_{0}\right\vert }%
{h}\,d(c,\left\{  a,b\right\}  )}\right)  =\mathcal{O}(\varepsilon
)\quad\forall t \label{Gamma_t}%
\end{equation}
where the definition (\ref{eps}) is taken into account. The corresponding
resonant state, given by the Green's function $G^{E(t)}(\cdot,c)$, will be
simply denoted by $G(t)$\thinspace.

The condition $\theta_{0}=ih^{N_{0}}$ in (h1) allows to control the
perturbation introduced by the interface conditions: namely, the distance
between $E(t)$ and the corresponding resonant level for the unperturbed model
is bounded by: $\mathcal{O}(\theta_{0}e^{-\frac{\left\vert \alpha
_{t}\right\vert }{h}\,d(c,\left\{  a,b\right\}  )})$, according to
(\ref{eigenvalue_exp}). A suitable choice of the parameter $d_{0}$ in
(h2)-(h3), ensures that: $E_{R}^{\frac{1}{2}}(t)\subset$supp $g\subset\left(
0,V_{0}\right)  $ definitely holds as $h\rightarrow0$.

The conditions (h1)-(h4) also provide with a well-posed functional analytical
framework for the study of the adiabatic problem. According to the result of
Proposition 3.7(d)) in \cite{FMN2}, the Hamiltonian $-\frac{i}{\varepsilon
}H_{\theta_{0},\alpha(t)}^{h}(\theta_{0})$, with $\alpha_{t}\in C^{2}\left(
(0,T),\mathbb{R}\right)  $, generates a dynamical system of contractions,
$S^{\varepsilon}(t,s)$, $t\geq s$, defined by the equation%
\begin{equation}
i\varepsilon\partial_{t}S^{\varepsilon}(t,s)=H_{\theta_{0},\alpha(t)}%
^{h}(\theta_{0})S^{\varepsilon}(t,s),\qquad S^{\varepsilon}%
(s,s)=Id\,,\label{S_t,s}%
\end{equation}
which preserves the domains, $S^{\varepsilon}(t,s)D(H_{\theta_{0},\alpha
(s)}^{h})\subset D(H_{\theta_{0},\alpha(t)}^{h})$ for $t\geq s$\thinspace. An
adiabatic theorem, for arbitrarily large time scales $\varepsilon
=e^{-\frac{\tau}{h}}$, has been proved to hold for a wide class of
Hamiltonians with interface conditions and exterior complex dilations,
including the case of $H_{\theta_{0},\alpha(t)}^{h}(\theta_{0})$ (Theorem~7.1
in \cite{FMN2}). To fix this point, consider the adiabatic evolution of the
initial resonant state $G(0)$; our problem is%
\begin{equation}
\medskip i\varepsilon\partial_{t}u=H_{\theta_{0},\alpha(t)}^{h}(\theta
_{0})u\,,\qquad u_{t=0}=G(0)\,,\label{Adiab_eq}%
\end{equation}
where $\varepsilon$ is fixed to the exponentially small scale $\varepsilon
=e^{-\frac{\left\vert \alpha_{0}\right\vert }{h}d\left(  c,\left\{
a,b\right\}  \right)  }$. Let $\mathcal{G}_{h}(E(t))$ denotes the set%
\[
\mathcal{G}_{h}(E(t))=\mathcal{\tilde{G}}_{h}(E(t))\backslash\left\{
z\in\mathbb{C},\ d(z,E(t))\geq\frac{h^{N_{0}}}{C}\right\}
\]
with $\mathcal{\tilde{G}}_{h}(\cdot)$ defined by (\ref{G_h}). For $C$ suitably
large, this forms a non-empty subset of $\mathcal{G}_{h}(E(t))$ where we can
define the normalized non-selfadjoint projector on $G(t)$ as%
\begin{equation}
P(t)=\frac{1}{2\pi i}\int_{\gamma^{h}(t)}\left(  z-H_{\theta_{0},\alpha
(t)}^{h}(\theta_{0})\right)  ^{-1}dz\label{P_t}%
\end{equation}
being $\gamma^{h}(t)$ a smooth curve in $\mathcal{G}_{h}(E(t))$ simply
connected to $E(t)$. With this notation, the result of \cite{FMN2} rephrases
as follows%
\begin{equation}
\sup_{t\in\left[  0,T\right]  }\left\vert S^{\varepsilon}(t,s)G(0)-\phi
_{t}\right\vert _{L^{2}(\mathbb{R})}\leq\,\mathcal{\tilde{O}}(\varepsilon
)\,\left\vert G(0)\right\vert _{L^{2}(\mathbb{R})}\label{adiabatic_est}%
\end{equation}%
\begin{equation}
\phi_{t}=\mu(t)e^{-\frac{i}{\varepsilon}\int_{0}^{t}E(s)\,ds}G(t)\,,\qquad
\text{and: }\left\{
\begin{array}
[c]{l}%
\medskip\dot{\mu}(t)\left\vert G(t)\right\vert _{L^{2}(\mathbb{R})}^{2}%
=-\mu(t)\left\langle G(t),P(t)\partial_{t}G(t)\right\rangle \\
\mu(0)=1
\end{array}
\right.  \label{Phi_t}%
\end{equation}
The coefficient at the r.h.s. of the equation can be made explicit according
to $P(t)=\frac{\left\langle G^{\ast}(t),\cdot\right\rangle G(t)}{\left\langle
G^{\ast}(t),G(t)\right\rangle }$, where $G^{\ast}(t)=G^{E^{\ast}(t)}$ is the
anti-resonant function. Using the inequalities (\ref{Green_uplow}) and
(\ref{Green_der_est}), we have: $G(t)-G^{\ast}(t)=\mathcal{\tilde{O}}\left(
E(t)-E^{\ast}(t)\right)  =\mathcal{\tilde{O}}\left(  \varepsilon\right)  $ in
$L^{2}$, and%
\begin{multline*}
\left\langle G(t),P(t)\partial_{t}G(t)\right\rangle _{L^{2}(\mathbb{R})}%
=\frac{\left\vert G(t)\right\vert _{L^{2}(\mathbb{R})}^{2}}{\left\langle
G^{\ast}(t),G(t)\right\rangle _{L^{2}(\mathbb{R})}}\,\left\langle G^{\ast
}(t),\partial_{t}G(t)\right\rangle _{L^{2}(\mathbb{R})}\\
=\frac{\left\vert G(t)\right\vert _{L^{2}(\mathbb{R})}^{2}}{\left\vert
G(t)\right\vert _{L^{2}(\mathbb{R})}^{2}+\mathcal{\tilde{O}}\left(
\varepsilon\right)  }\,\left(  \left\langle G(t),\partial_{t}G(t)\right\rangle
_{L^{2}(\mathbb{R})}+\mathcal{\tilde{O}}\left(  \varepsilon\right)  \right)
=\left\langle G(t),\partial_{t}G(t)\right\rangle _{L^{2}(\mathbb{R}%
)}+\mathcal{\tilde{O}}\left(  \varepsilon\right)
\end{multline*}
This provides with an expansion for $\mu(t)$
\begin{equation}
\mu(t)=e^{-\int_{0}^{t}\frac{\left\langle G,\partial_{s}G\right\rangle
}{\left\vert G(s)\right\vert _{2}^{2}}ds}+\mathcal{\tilde{O}}\left(
\varepsilon\right)  \label{Mu_t}%
\end{equation}

\subsection{A decomposition of $A_{\theta_{0}}(t)$}

We shall use a decomposition in the same spirit of the one proposed in
\cite{JoPrSj} and \cite{PrSj} with additional specific information given by
our specific model. For $\theta=\theta_{0}$, the Cauchy problem (\ref{A_def2})
writes as%
\begin{equation}
\left\{
\begin{array}
[c]{l}%
\medskip i\varepsilon\partial_{t}u_{\theta_{0}}(k,\cdot,t)=H_{\theta
_{0},\alpha(t)}^{h}(\theta_{0})u_{\theta_{0}}(k,\cdot,t)\,,\\
u_{t=0}=U_{\theta_{0}}\psi_{-}(k,\cdot,\alpha_{0})\,,
\end{array}
\right.  \label{gen_eigenf_evolution}%
\end{equation}
where, for a fixed $\alpha$, $U_{\theta_{0}}\psi_{-}(k,\cdot,\alpha)$ solves
the equation%
\begin{equation}
\left(  H_{\theta_{0},\alpha}^{h}(\theta_{0})-k^{2}\right)  U_{\theta_{0}}%
\psi_{-}(k,\cdot,\alpha)=0,\qquad H_{\theta_{0},\alpha}^{h}(\theta_{0})u\in
L_{loc}^{2}\,. \label{gen_eigenf_def}%
\end{equation}
For time dependent $\alpha$, the following representation holds%
\begin{equation}
U_{\theta_{0}}\psi_{-}(k,\cdot,\alpha(t))=U_{\theta_{0}}\tilde{\psi}%
_{-}(k,\cdot)+C(k,t)G^{k^{2}}\,, \label{gen_eigenf_def1}%
\end{equation}
where $\tilde{\psi}_{-}(k,\cdot)$ are the incoming scattering states of the
unperturbed Hamiltonian (solving (\ref{gen_eigenf})), the coefficient $C(k,t)$
is defined according to%
\begin{equation}
C(k,t)=-\frac{h\alpha_{t}\,\tilde{\psi}_{-}(k,c)}{1+h\alpha_{t}\,G^{k^{2}}%
(c)}\,, \label{C_kt}%
\end{equation}
while $G^{k^{2}}$ is explicitely given in (\ref{Green_ext})-(\ref{Green_in}).
A possible decomposition of the solution of (\ref{gen_eigenf_evolution}) is%
\begin{equation}
u_{\theta_{0}}(k,\cdot,t)=e^{-i\frac{t}{\varepsilon}k^{2}}U_{\theta_{0}}%
\psi_{-}(k,\cdot,\alpha(t))+R(t)\,. \label{decomp1}%
\end{equation}
Denoting $\varphi_{t}=e^{-i\frac{t}{\varepsilon}k^{2}}U_{\theta_{0}}\psi
_{-}(k,\cdot,\alpha(t))$, we have%
\begin{equation}
\left\{
\begin{array}
[c]{l}%
\medskip i\varepsilon\,\partial_{t}\varphi_{t}=k^{2}\varphi_{t}+i\varepsilon
\,e^{-i\frac{t}{\varepsilon}k^{2}}\dot{C}(k,t)G^{k^{2}}=H_{\theta_{0}%
,\alpha(t)}^{h}(\theta_{0})\varphi_{t}+i\varepsilon\,e^{-i\frac{t}%
{\varepsilon}k^{2}}\dot{C}(k,t)G^{k^{2}}\,,\\
\varphi_{0}=U_{\theta_{0}}\psi_{-}(k,\cdot,\alpha_{0})\,.
\end{array}
\right.  \label{small_phi_eq}%
\end{equation}
As it follows from (\ref{A_def2}) and (\ref{small_phi_eq}), the remainder
$R(t)=u_{\theta}(k,\cdot,t)-\varphi_{t}$ solves the Cauchy problem%
\begin{equation}
\left\{
\begin{array}
[c]{l}%
\medskip i\varepsilon\partial_{t}R(t)=H_{\theta_{0},\alpha(t)}^{h}(\theta
_{0})R(t)+i\varepsilon\,e^{-i\frac{t}{\varepsilon}k^{2}}\dot{C}(k,t)G^{k^{2}%
}\,,\\
R(0)=0\,,
\end{array}
\right.
\end{equation}
and its explicit form is%
\begin{equation}
R(t)=-\int_{0}^{t}S^{\varepsilon}(t,s)\,e^{-i\frac{s}{\varepsilon}k^{2}}%
\dot{C}(k,s)G^{k^{2}}ds\,, \label{Remainder}%
\end{equation}
where $S^{\varepsilon}(t,s)$ is the dynamical system associated with
$-\frac{i}{\varepsilon}H_{\theta_{0},\alpha(t)}^{h}(\theta_{0})$. Making use
of (\ref{decomp1}) and (\ref{Remainder}), the time evolution $u_{\theta_{0}%
}(k,\cdot,t)$ further decomposes in the sum%
\begin{equation}
u_{\theta_{0}}(k,\cdot,t)=\sum_{j=1}^{4}\psi_{j}(k,\cdot,t)\,,
\label{decomp1_4}%
\end{equation}
with%
\begin{equation}
\psi_{1}(k,\cdot,t)=e^{-i\frac{t}{\varepsilon}k^{2}}\left[  U_{\theta_{0}%
}\tilde{\psi}_{-}(k,\cdot)+C(k,t)\left(  G^{k^{2}}-G(t)\right)  \right]  \,,
\label{psi_1}%
\end{equation}%
\begin{equation}
\psi_{2}(k,\cdot,t)=-\int_{0}^{t}S^{\varepsilon}(t,s)\,\dot{C}%
(k,s)\,e^{-i\frac{s}{\varepsilon}k^{2}}\,\left(  G^{k^{2}}-G(s)\right)  ds\,,
\label{psi_2}%
\end{equation}%
\begin{equation}
\psi_{3}(k,\cdot,t)=-\int_{0}^{t}\dot{C}(k,s)\,e^{-i\frac{s}{\varepsilon}%
k^{2}}\,\left(  S^{\varepsilon}(t,s)G(s)-\mu(t)e^{-\frac{i}{\varepsilon}%
\int_{s}^{t}E(\sigma)\,d\sigma}G(t)\right)  ds\,, \label{psi_3}%
\end{equation}%
\begin{gather}
\medskip\psi_{4}(k,\cdot,t)=e^{-i\frac{t}{\varepsilon}k^{2}}\left[
C(k,t)-\int_{0}^{t}\dot{C}(k,s)\,e^{-\frac{i}{\varepsilon}\int_{s}^{t}\left(
E(\sigma)-k^{2}\right)  \,d\sigma}ds\right]  \mu(t)G(t)\qquad\nonumber\\
\qquad\qquad\qquad\qquad\qquad\qquad\qquad\qquad+e^{-i\frac{t}{\varepsilon
}k^{2}}\left(  1-\mu(t)\right)  C(k,t)G(t)\,. \label{psi_4}%
\end{gather}
The variable $A_{\theta_{0}}(t)$, associated with $u_{\theta_{0}}(k,\cdot,t)$,
now writes as%
\begin{equation}
A_{\theta_{0}}(t)=\sum_{j,j^{\prime}=1}^{4}\int\frac{dk}{2\pi h}%
\,g(k)\,\left\langle \chi\psi_{j}(k,\cdot,t),\psi_{j^{\prime}}(k,\cdot
,t)\right\rangle _{L^{2}(\mathbb{R})}\,. \label{A_decomp}%
\end{equation}

In order to get adiabatic estimates for the contributions to (\ref{A_decomp}),
we need accurate asymptotic expansions for the quantities involved in these
computations, including: $\tilde{\psi}_{-}(k,\cdot)$, and the integrals of
$\left\vert C(k,t)\right\vert ^{2}$. To this aim we introduce the following
technical lemma.

\begin{lemma}
\label{Lemma_3}In the assumptions (h1)-(h4), let $E\in\mathbb{R}$,
$E^{\frac{1}{2}}\in$supp$g$ and $\lambda_{t}=\lim_{h\rightarrow0}E(t)$; the
solutions to (\ref{gen_eigenf}) for $k^{2}=E$ fulfill the conditions%
\begin{equation}
\left\vert \tilde{\psi}_{-}(E^{\frac{1}{2}},c)\right\vert ^{2}=e^{-\frac
{\left\vert \alpha_{t}\right\vert }{h}(c-a)}\mathcal{O}(1)\,;\qquad\left\vert
\tilde{\psi}_{-}(-E^{\frac{1}{2}},c)\right\vert ^{2}=e^{-\frac{\left\vert
\alpha_{t}\right\vert }{h}(b-c)}\mathcal{O}(1)\,, \label{gen_eigenf^2_exp0}%
\end{equation}
for all $c\in\left(  a,b\right)  $. In particular, for $\left\vert
E-\lambda_{t}\right\vert <C\varepsilon$ and $d(c,\left\{  a,b\right\}  )=c-a$,
the first of (\ref{gen_eigenf^2_exp0}) is explicitly%
\begin{equation}
\left\vert \tilde{\psi}_{-}(E^{\frac{1}{2}},c)\right\vert ^{2}=2\lambda
_{t}^{\frac{1}{2}}\left\vert \alpha_{t}\right\vert \frac{\Gamma_{t}%
}{\left\vert M(\lambda_{t},\lambda_{t})\right\vert ^{2}}\left(  1+\mathcal{O}%
\left(  \left\vert \theta_{0}\right\vert \right)  \right)  +o(\varepsilon)\,,
\label{gen_eigenf^2_exp1}%
\end{equation}
$M(E_{1},E_{2})$, $\Gamma_{t}$ and $\varepsilon$ being defined according to
(\ref{M_E_def}), (\ref{Res_imm}) and (\ref{eps}) respectively. The function
$\tilde{\psi}_{-}(E^{\frac{1}{2}},c)$ holomorphically extends to the complex
neighbourhood $\mathbb{C}\cap\left\{  \left\vert z-E_{res}^{h}\right\vert
\leq\frac{h}{d_{0}}\right\}  $ where the representation%
\begin{equation}
\tilde{\psi}_{-}(E^{\frac{1}{2}},c)\tilde{\psi}_{-}^{\ast}(\left(  E^{\ast
}\right)  ^{\frac{1}{2}},c)=\,e^{-\frac{\left\vert \alpha\right\vert }%
{h}(c-a)}\,\mathcal{F}^{h,\theta_{0}}(E) \label{gen_eigenf^2_exp2}%
\end{equation}
holds, being $\mathcal{F}^{h,\theta_{0}}(\cdot)$ an holomorphic family
uniformly bounded w.r.t. $h$ and $\theta_{0}$.
\end{lemma}

\begin{preuve}
In (\ref{gen_eigenf_in}) the explicit form of $\tilde{\psi}_{-}(k,\cdot)$,
$k>0$, is given. For energies $k^{2}$ placed below the barrier level $V_{0}$,
the decreasing behaviour of the terms $e^{-i\frac{\Lambda_{z}}{h}\,l}$,
$e^{-i\frac{\Lambda_{z}}{h}\,(c-b)}$ w.r.t. $h$ allow to write%
\begin{equation}
\tilde{\psi}_{-}(k,c)=e^{-i\frac{\Lambda_{k^{2}}}{h}(c-a)}\,\mathcal{P}%
^{h}\left(  k^{2},\theta_{0}\right)  \label{gen_eigenf_exp}%
\end{equation}
with $\Lambda_{z}$ defined as in (\ref{small_gamma}), $\mathcal{P}^{h}\left(
z,\theta_{0}\right)  $ an holomorphic map w.r.t. both variables, provided that
$\left\vert z-E_{res}^{h}\right\vert <\frac{h}{d_{0}}$ and $\left\vert
\theta_{0}\right\vert ,h$ are small enough. The first part of
(\ref{gen_eigenf^2_exp0}) follow by using (\ref{gen_eigenf_exp}) with:
$k^{2}=E=\lambda_{t}+\mathcal{O}(h)$. The second part of
(\ref{gen_eigenf^2_exp0}) can be carried out by a similar direct computation.
For $h\rightarrow0$, the asymptotic behaviour of $\tilde{\psi}_{-}(k,c)$ is
determined by the factor $e^{-i\frac{\Lambda_{k^{2}}}{h}(c-a)}$. In the
complex neighbourhood $\left\vert z-E_{res}^{h}\right\vert <\frac{h}{d_{0}}$,
where $E_{res}^{h}=V_{0}-\frac{\alpha^{2}}{4}+\mathcal{O}\left(
e^{-\frac{\left\vert \alpha_{t}\right\vert }{h}\,d(c,\left\{  a,b\right\}
)}\right)  $, the function $\Lambda_{z}$ is analytic and the relation:
$e^{-i\frac{\Lambda_{z}}{h}(c-a)}=e^{-\frac{-\left\vert \alpha\right\vert
}{2h}(c-a)}\mathcal{R}^{h}(z)$ holds being $\mathcal{R}^{h}(\cdot)$ an
analytic family uniformly bounded w.r.t. $h$. A similar identity holds for
$\tilde{\psi}_{-}(k,c)$, once (\ref{gen_eigenf_exp}) is taken into account.
The representation (\ref{gen_eigenf^2_exp2}) is a direct consequence of this relation.

Next we use the notation $\tilde{\psi}_{-,\theta_{0}}(k,\cdot)$ and
$G_{\theta_{0}}^{E}(\cdot,c)$ to point out the dependence of scattering states
and Green's functions on the interface conditions of the Hamiltonian. When
$\theta_{0}=0$, $H_{0,0}^{h}(0)$ is a selfadjoint operator with purely
absolutely continuous spectrum. The Stone's formula yields in this case%
\begin{multline*}
\int_{0}^{+\infty}\frac{dk}{2\pi h}f(k^{2})\left[  \left\vert \tilde{\psi
}_{-,0}(k,c)\right\vert ^{2}+\left\vert \tilde{\psi}_{-,0}(-k,c)\right\vert
^{2}\right] \\
=\frac{1}{2\pi i}\lim_{\delta\rightarrow0}\int_{0}^{+\infty}%
dE\,f(E)\left\langle \delta_{c},\left[  \left(  H_{0,0}^{h}(0)-E+i\delta
\right)  ^{-1}-\left(  H_{0,0}^{h}(0)-E-i\delta\right)  ^{-1}\right]
\delta_{c}\right\rangle _{H^{-1},H^{1}}\\
=\frac{1}{\pi}\lim_{\delta\rightarrow0}\int dE\,f(E)\operatorname{Im}%
G_{0}^{E-i\delta}(c,c)=\frac{1}{\pi}\lim_{\delta\rightarrow0}\int
dk\,2\left\vert k\right\vert f(k^{2})\operatorname{Im}G_{0}^{k^{2}-i\delta
}(c,c)\,.
\end{multline*}
for continuous $f$. This leads to%
\begin{equation}
\left\vert \tilde{\psi}_{-,0}(k,c)\right\vert ^{2}+\left\vert \tilde{\psi
}_{-,0}(-k,c)\right\vert ^{2}=4h\left\vert k\right\vert \,\operatorname{Im}%
G_{0}^{k^{2}-i0}(c,c)\,. \label{gen_eigenf^2_exp3}%
\end{equation}
For $\left\vert E-\lambda_{t}\right\vert <C\varepsilon$,
(\ref{gen_eigenf^2_exp0}) implies: $\left\vert \tilde{\psi}_{-,0}(-E^{\frac
{1}{2}},c)\right\vert ^{2}=\mathcal{O}(e^{-\frac{\left\vert \alpha
_{t}\right\vert }{h}(b-c)})$, and due to the assumption $d(c,\left\{
a,b\right\}  )=c-a$, we get%
\begin{equation}
\left\vert \tilde{\psi}_{-,0}(E^{\frac{1}{2}},c)\right\vert ^{2}=4hE^{\frac
{1}{2}}\,\operatorname{Im}G_{0}^{E-i0}(c,c)+o(\varepsilon)\,.
\end{equation}
The relation (\ref{gen_eigenf^2_exp1}), for $\theta_{0}=0$, follows by using
(\ref{exp_1.1}) and (\ref{M_E}) to express $\operatorname{Im}G_{0}%
^{E-i0}(c,c)$ as a function of $M(E,E(t))$, and expanding for $E=\lambda
_{t}-i\Gamma_{t}+o(\varepsilon)$. The general case is recovered by noticing
that (\ref{gen_eigenf_exp}), and the correspondent expression for $\tilde
{\psi}_{-,\theta_{0}}(-k,\cdot)$, imply%
\begin{equation}
\left\vert \tilde{\psi}_{-,\theta_{0}}(k,c)\right\vert ^{2}+\left\vert
\tilde{\psi}_{-,\theta_{0}}(-k,c)\right\vert ^{2}=\left(  \left\vert
\tilde{\psi}_{-,0}(k,c)\right\vert ^{2}+\left\vert \tilde{\psi}_{-,0}%
(-k,c)\right\vert ^{2}\right)  \left(  1+\mathcal{O}\left(  \left\vert
\theta_{0}\right\vert \right)  \right)  \,. \label{gen_eigenf_small_theta}%
\end{equation}
(Actually (\ref{gen_eigenf_small_theta}) could be also recovered, with the
less efficient bound $\mathcal{O}\left(  h^{-1}\left\vert \theta
_{0}\right\vert \right)  $, from Propositions 4.5 in \cite{FMN2}). Thus, for
$\left\vert \theta_{0}\right\vert <<1$ (we refer to the assumption (h1)),
(\ref{gen_eigenf^2_exp3}) writes as%
\begin{equation}
\left\vert \tilde{\psi}_{-,\theta_{0}}(k,c)\right\vert ^{2}+\left\vert
\tilde{\psi}_{-,\theta_{0}}(-k,c)\right\vert ^{2}=4h\left\vert k\right\vert
\,\operatorname{Im}G_{0}^{k^{2}-i0}(c,c)\left(  1+\mathcal{O}\left(
\left\vert \theta_{0}\right\vert \right)  \right)  \,.
\end{equation}
Proceeding as before, we obtain (\ref{gen_eigenf^2_exp1}).
\end{preuve}

Next computations involve the use of small-$h$ expansions of the coefficients
$C(k,t)$ and $\dot{C}(k,t)$. Using the definition (\ref{C_kt}) and the
relation (\ref{exp_1.1}), leads%
\begin{equation}
C(k,t)=-\frac{h\alpha_{t}\,\tilde{\psi}_{-}(k,c)\,M(k^{2},E(t))}{k^{2}%
-E(t)}\,. \label{C_kt_exp}%
\end{equation}
The derivative $\dot{C}(k,s)$ is explicitly given by%
\begin{equation}
\dot{C}(k,t)=\frac{h\,\dot{\alpha}_{t}\,\tilde{\psi}_{-}(k,c)}{\left(
1+h\alpha_{t}\,G^{k^{2}}(c)\right)  ^{2}}\left(  2h\alpha_{t}\,G^{k^{2}%
}(c)-1\right)  =\frac{h\,\dot{\alpha}_{t}\,\tilde{\psi}_{-}(k,c)\,M(k^{2}%
,E(t))}{\left(  k^{2}-E(t)\right)  ^{2}}\mathcal{O}(1) \label{C_kt_dot}%
\end{equation}
for all $k\in$supp $g$, and $t\in\left[  0,T\right]  $. The relations
(\ref{gen_eigenf^2_exp0}) and (\ref{gen_eigenf^2_exp1}), allows to identify
$\left\vert C(k,t)\right\vert ^{2}$ with a Lorentzian function on supp $g$,
with scale parameter given by $\Gamma_{t}$. In particular, for: $d(c,\left\{
a,b\right\}  )=c-a$, it holds%
\begin{equation}
\int\frac{dk}{2\pi h}\,g(k)\,\left\vert C(k,t)\right\vert ^{2}=\frac
{h\left\vert \alpha_{t}\right\vert ^{3}}{2}\,g\left(  \lambda_{t}^{\frac{1}%
{2}}\right)  \left(  1+\mathcal{O}\left(  \left\vert \theta_{0}\right\vert
\right)  \right)  +o(\varepsilon)\,, \label{Int_1}%
\end{equation}
while, for $d(c,\left\{  a,b\right\}  )=b-c$, we have%
\begin{equation}
\int\frac{dk}{2\pi h}\,g(k)\,\left\vert C(k,t)\right\vert ^{2}=\mathcal{O}%
\left(  e^{-\frac{\beta}{h}}\right)  \,, \label{Int_1.1}%
\end{equation}
with positive $\beta=\frac{\left\vert \alpha_{t}\right\vert }{h}(c-a-(b-c))$.
Both the above expansions follow by using the dilation $y=\frac{k^{2}%
-E(t)}{\Gamma_{t}}$ and taking the limit of the resulting integral as
$h\rightarrow0$. With a similar computation we also have%
\begin{equation}
\int dk\,g(k)\,\frac{1}{\left\vert k^{2}-E(t)\right\vert }=\mathcal{O}\left(
\frac{1}{h}\right)  \,. \label{Int_2}%
\end{equation}

\begin{lemma}
\label{Lemma_diag}In the assumptions (h1)-(h4), the estimates:%
\begin{equation}
\left\vert \int\frac{dk}{2\pi h}\,g(k)\,\left\langle \chi\,\psi_{j}%
(k,\cdot,t),\psi_{j}(k,\cdot,t)\right\rangle _{L^{2}(\mathbb{R})}\right\vert
=\mathcal{\tilde{O}}(\varepsilon^{\frac{1}{J+\delta}}) \label{diag_bounds}%
\end{equation}
hold with: $j=1,2,3$.
\end{lemma}

\begin{preuve}
For $j=1$, this product develops in the sum%
\begin{multline*}
\int\frac{dk}{2\pi h}\,g(k)\,\left\langle \chi\,U_{\theta_{0}}\tilde{\psi}%
_{-}(k,\cdot),U_{\theta_{0}}\tilde{\psi}_{-}(k,\cdot)\right\rangle
_{L^{2}(\mathbb{R})}\\[0.3cm]
+\int\frac{dk}{2\pi h}\,g(k)\,\left\langle \chi\,C(k,t)\left(  G^{k^{2}%
}-G(t)\right)  ,C(k,t)\left(  G^{k^{2}}-G(t)\right)  \right\rangle
_{L^{2}(\mathbb{R})}+\\
2\operatorname{Re}\int\frac{dk}{2\pi h}\,g(k)\,\left\langle \chi
\,U_{\theta_{0}}\tilde{\psi}_{-}(k,\cdot),C(k,t)\left(  G^{k^{2}}-G(t)\right)
\right\rangle _{L^{2}(\mathbb{R})}\,.
\end{multline*}
Using the exponential decreasing behaviour of $\tilde{\psi}_{-}(k,\cdot)$ on
the supp $\chi$ (see Lemma \ref{Lemma_3}), the first contribution is estimated
by%
\begin{equation}
\medskip\left\vert \int\frac{dk}{2\pi h}\,g(k)\,\left\langle \chi
\,U_{\theta_{0}}\tilde{\psi}_{-}(k,\cdot),U_{\theta_{0}}\tilde{\psi}%
_{-}(k,\cdot)\right\rangle _{L^{2}(\mathbb{R})}\right\vert =\mathcal{\tilde
{O}}\left(  e^{-2\frac{\left\vert \alpha_{0}\right\vert }{h}(c-a)}\right)  \,.
\label{diag1.1_bound}%
\end{equation}
For the second term, the definition of $C(k,t)$ (see (\ref{C_kt_exp})), the
equivalence (\ref{gen_eigenf^2_exp0}) and the first inequality in
(\ref{Green_der_est}) lead to%
\begin{align}
&  \medskip\left\vert \int\frac{dk}{2\pi h}\,g(k)\,\left\vert
C(k,t)\right\vert ^{2}\left\langle \chi\,\left(  G^{k^{2}}-G(t)\right)
,\left(  G^{k^{2}}-G(t)\right)  \right\rangle _{L^{2}(\mathbb{R})}\right\vert
\nonumber\\[0.3cm]
&  \leq C\int\frac{dk}{2\pi h}\,\left\vert g(k)\right\vert \,\left\vert
C(k,t)\right\vert ^{2}\left\Vert G^{k^{2}}-G(t)\right\Vert _{L^{2}%
(\mathbb{R})}=\mathcal{\tilde{O}}\left(  e^{-\frac{\left\vert \alpha
_{t}\right\vert }{h}(c-a)}\right)  \,. \label{diag1.2_bound}%
\end{align}
The last contribution is a crossing term; it is estimated in terms of the
previous ones by using the H\"{o}lder inequality in $L^{2}(\mathbb{R}^{2})$%
\begin{equation}
\left\vert \int\frac{dk}{2\pi h}\,g(k)\,\left\langle \chi\,U_{\theta_{0}%
}\tilde{\psi}_{-}(k,\cdot),C(k,t)\left(  G^{k^{2}}-G(t)\right)  \right\rangle
_{L^{2}(\mathbb{R})}\right\vert \leq\mathcal{\tilde{O}}(\varepsilon)\,.
\label{diag1.3_bound}%
\end{equation}

For $j=2$, the integral is%
\begin{equation}
\int\frac{dk}{2\pi h}\,g(k)\,\left\langle \chi\psi_{2}(k,\cdot,t),\psi
_{2^{\prime}}(k,\cdot,t)\right\rangle _{L^{2}(\mathbb{R})}=\int\frac{dk}{2\pi
h}\int_{0}^{t}ds_{1}\int_{0}^{t}ds_{2}\,f(k,s_{1},s_{2})\,,
\end{equation}
with%
\begin{align}
f(k,s_{1},s_{2})  &  =g(k)\,\dot{C}(k,s_{1})\dot{C}^{\ast}(k,s_{2}%
)\,e^{-i\frac{\left(  s_{1}-s_{2}\right)  }{\varepsilon}k^{2}}\,\times
\nonumber\\
&  \times\left\langle \chi\,S^{\varepsilon}(t,s_{1})\left(  G^{k^{2}}%
-G(s_{1})\right)  ,S^{\varepsilon}(t,s_{2})\left(  G^{k^{2}}-G(s_{s})\right)
\right\rangle _{L^{2}(\mathbb{R})}\,.
\end{align}
Since $f(k,s_{1},s_{2})=f^{\ast}(k,s_{2},s_{1})$, this integral writes as%
\begin{equation}
\int\frac{dk}{2\pi h}\,g(k)\,\left\langle \chi\psi_{2}(k,\cdot,t),\psi
_{2^{\prime}}(k,\cdot,t)\right\rangle _{L^{2}(\mathbb{R})}=2\operatorname{Re}%
\int\frac{dk}{2\pi h}\int_{0}^{t}ds_{1}\int_{0}^{s_{1}}ds_{2}\,f(k,s_{1}%
,s_{2})\,. \label{diag_2}%
\end{equation}
The first inequality of (\ref{Green_der_est}), leads: $\left\Vert
S^{\varepsilon}(t,s)\left(  G^{k^{2}}-G(s)\right)  \right\Vert _{L^{2}%
(\mathbb{R})}\leq\left\vert k^{2}-E(s)\right\vert \,\mathcal{\tilde{O}}\left(
\varepsilon^{0}\right)  $. Then, according to the definitions (\ref{C_kt_exp}%
)-(\ref{C_kt_dot}), we find
\[
\left\vert f(k,s_{1},s_{2})\right\vert \leq\mathcal{\tilde{O}}\left(
\varepsilon^{0}\right)  \left\vert g(k)\right\vert \left\vert C(k,s_{1}%
)C^{\ast}(k,s_{2})\right\vert \,,
\]
and%
\begin{equation}
\left\vert \int\frac{dk}{2\pi h}\int_{0}^{t}ds_{1}\int_{0}^{s_{1}}%
ds_{2}\,f(k,s_{1},s_{2})\right\vert \leq\mathcal{\tilde{O}}(\varepsilon
^{0})\int_{0}^{t}ds_{1}\int_{0}^{s_{1}}ds_{2}\int dk\left\vert \frac
{g(k)}{2\pi h}\right\vert \left\vert C(k,s_{1})C^{\ast}(k,s_{2})\right\vert
\,. \label{diag_2_est0}%
\end{equation}
The integral over $k$ admits two independent estimates:\medskip\newline$1)$
Use the Cauchy-Schwarz inequality to write%
\begin{gather}
\int dk\left\vert \frac{g(k)}{2\pi h}\right\vert \left\vert C(k,s_{1})C^{\ast
}(k,s_{2})\right\vert \qquad\qquad\qquad\qquad\qquad\qquad\qquad\qquad
\qquad\qquad\nonumber\\
\qquad\qquad\leq\left(  \int\frac{dk}{2\pi h}\,g(k)\,\left\vert C(k,s_{1}%
)\right\vert ^{2}\right)  ^{\frac{1}{2}}\left(  \int\frac{dk}{2\pi
h}\,g(k)\,\left\vert C(k,s_{2})\right\vert ^{2}\right)  ^{\frac{1}{2}%
}=\mathcal{O}(1)\,. \label{diag_2_est1}%
\end{gather}
$2)$ Use (\ref{C_kt_exp}) and the relation%
\[
\frac{\left\vert E(s_{1})-E(s_{2})\right\vert }{\left\vert k^{2}%
-E(s_{1})\right\vert \,\left\vert k^{2}-E(s_{2})\right\vert }=\left\vert
\frac{1}{k^{2}-E(s_{1})}-\frac{1}{k^{2}-E(s_{2})}\right\vert
\]
to write%
\[
\int dk\left\vert \frac{g(k)}{2\pi h}\right\vert \left\vert C(k,s_{1})C^{\ast
}(k,s_{2})\right\vert \leq\frac{\mathcal{\tilde{O}}(\varepsilon^{0}%
)}{\left\vert E(s_{1})-E(s_{2})\right\vert }\left(  \int\limits_{\text{supp
}g}dk\frac{\left\vert \tilde{\psi}_{-}(k,c)\right\vert ^{2}}{\left\vert
k^{2}-E(s_{1})\right\vert }+\int\limits_{\text{supp }g}dk\frac{\left\vert
\tilde{\psi}_{-}(k,c)\right\vert ^{2}}{\left\vert k^{2}-E(s_{2})\right\vert
}\right)  \,.
\]
From (\ref{Int_2}) and (\ref{gen_eigenf^2_exp0}), it follows%
\begin{equation}
\int dk\left\vert \frac{g(k)}{2\pi h}\right\vert \left\vert C(k,s_{1})C^{\ast
}(k,s_{2})\right\vert \leq\frac{\mathcal{\tilde{O}}\left(  e^{-\frac
{\left\vert \alpha_{t}\right\vert }{h}(c-a)}\right)  }{\left\vert
E(s_{1})-E(s_{2})\right\vert }\,. \label{diag_2_est2}%
\end{equation}

Interpolating between (\ref{diag_2_est1}) and (\ref{diag_2_est2}) yields%
\[
\int dk\left\vert \frac{g(k)}{2\pi h}\right\vert \left\vert C(k,s_{1})C^{\ast
}(k,s_{2})\right\vert \leq\frac{\mathcal{\tilde{O}}\left(  e^{-\frac
{\left\vert \alpha_{t}\right\vert }{n\,h}(c-a)}\right)  }{\left\vert
E(s_{1})-E(s_{2})\right\vert ^{\frac{1}{n}}}\leq\frac{\mathcal{\tilde{O}%
}\left(  e^{-\frac{\left\vert \alpha_{t}\right\vert }{n\,h}(c-a)}\right)
}{\left\vert \alpha(s_{1})-\alpha(s_{2})\right\vert ^{\frac{1}{n}}}\,,
\]
where we use the lower bound: $\left\vert E(s_{1})-E(s_{2})\right\vert \geq
c_{0}\left\vert \alpha(s_{1})-\alpha(s_{2})\right\vert $ following from
(\ref{h2_real})-(\ref{h2_immaginary}) and the analyticity of the map
$\alpha\rightarrow E_{res}^{h}$ (see Proposition \ref{Prop_1}). Due to the
assumption (h2), $\left\vert \alpha(s_{1})-\alpha(s_{2})\right\vert
^{-\frac{1}{n}}$ is integrable on the triangle $\left\{  s_{1}\in\left[
0,T\right]  ,s_{2}\leq s_{1}\right\}  $ provided that $n\geq J+\delta$. This
leads to%
\[
\left\vert \int\frac{dk}{2\pi h}\,g(k)\,\left\langle \chi\psi_{2}%
(k,\cdot,t),\psi_{2^{\prime}}(k,\cdot,t)\right\rangle _{L^{2}(\mathbb{R}%
)}\right\vert =\mathcal{\tilde{O}}\left(  e^{-\frac{\left\vert \alpha
_{t}\right\vert }{n\,h}(c-a)}\right)
\]

For $j=3$, the integral is%
\begin{align*}
&  \int\frac{dk}{2\pi h}\,g(k)\,\left\langle \chi\psi_{3}(k,\cdot
,t),\psi_{3^{\prime}}(k,\cdot,t)\right\rangle _{L^{2}(\mathbb{R})}\\
&  =\int_{0}^{t}ds_{1}\int_{0}^{t}ds_{2}\int\frac{dk}{2\pi h}\,g(k)\,\dot
{C}(k,s_{1})\dot{C}^{\ast}(k,s_{2})\,\left\langle \chi\varphi(t,s_{1}%
),\varphi(t,s_{2})\right\rangle _{L^{2}(\mathbb{R})}\,,
\end{align*}
where%
\[
\varphi(t,s)=S^{\varepsilon}(t,s)G(s)-\mu(t)e^{-\frac{i}{\varepsilon}\int
_{s}^{t}E(\sigma)\,d\sigma}G(t)\,.
\]
Using (\ref{adiabatic_est}), (\ref{Phi_t}) and (\ref{Mu_t}), it follows:
$\left\vert \varphi(t,s)\right\vert _{L^{2}(\mathbb{R})}\leq\mathcal{\tilde
{O}}(\varepsilon)$ uniformly w.r.t. $t$ and $s$; then, proceeding as above, we
get%
\[
\left\vert \int\frac{dk}{2\pi h}\,g(k)\,\left\langle \chi\psi_{3}%
(k,\cdot,t),\psi_{3^{\prime}}(k,\cdot,t)\right\rangle _{L^{2}(\mathbb{R}%
)}\right\vert \leq\mathcal{\tilde{O}}(\varepsilon^{2})\int_{0}^{t}ds_{1}%
\int_{0}^{t}ds_{2}\int dk\left\vert \frac{g(k)}{2\pi h}\right\vert
\frac{\left\vert C(k,s_{1})C^{\ast}(k,s_{2})\right\vert }{\left\vert
k^{2}-E(s_{1})\right\vert \left\vert k^{2}-E(s_{2})\right\vert }\,.
\]
Since $\left\vert k^{2}-E(s)\right\vert ^{-1}\leq\frac{1}{\varepsilon}$ on
supp $g$, a similar inequality to the one considered in (\ref{diag_2_est0})
follows. We obtain%
\[
\left\vert \int\frac{dk}{2\pi h}\,g(k)\,\left\langle \chi\psi_{3}%
(k,\cdot,t),\psi_{3^{\prime}}(k,\cdot,t)\right\rangle _{L^{2}(\mathbb{R}%
)}\right\vert \leq\mathcal{\tilde{O}}\left(  e^{-\frac{\left\vert \alpha
_{t}\right\vert }{n\,h}(c-a)}\right)  \,.
\]

\end{preuve}

\subsection{The reduced equation.}

We consider the term%
\begin{equation}
\int\frac{dk}{2\pi h}\,g(k)\,\left\langle \chi\psi_{4}(k,\cdot,t),\psi
_{4^{\prime}}(k,\cdot,t)\right\rangle \,.\label{A_p}%
\end{equation}
Setting: $\psi_{4}(k,\cdot,t)=\varphi_{1}(k,\cdot,t)+\varphi_{2}(k,\cdot,t)$,%
\begin{align}
\bigskip\varphi_{1}(k,\cdot,t) &  =e^{-i\frac{t}{\varepsilon}k^{2}}\left[
C(k,t)-\int_{0}^{t}\dot{C}(k,s)\,e^{-\frac{i}{\varepsilon}\int_{s}^{t}\left(
E(\sigma)-k^{2}\right)  \,d\sigma}ds\right]  \mu(t)G(t)\,,\label{phi_1}\\
\varphi_{2}(k,\cdot,t) &  =e^{-i\frac{t}{\varepsilon}k^{2}}\left(
1-\mu(t)\right)  C(k,t)G(t)\,,\label{phi_2}%
\end{align}
and introducing the variables%
\begin{align}
\medskip a(t) &  =\int\frac{dk}{2\pi h}\,g(k)\,\left\langle \chi\varphi
_{1}(k,\cdot,t),\varphi_{1}(k,\cdot,t)\right\rangle \,,\label{a_reduc}\\
\medskip\mathcal{J}_{1}(t) &  =\int\frac{dk}{2\pi h}\,g(k)\,\left\langle
\chi\varphi_{2}(k,\cdot,t),\varphi_{2}(k,\cdot,t)\right\rangle \,,\label{J1}\\
\mathcal{J}_{2}(t) &  =2\operatorname{Re}\int\frac{dk}{2\pi h}%
\,g(k)\,\left\langle \chi\varphi_{1}(k,\cdot,t),\varphi_{2}(k,\cdot
,t)\right\rangle \,,\label{J2}%
\end{align}
it becomes%
\begin{equation}
\int\frac{dk}{2\pi h}\,g(k)\,\left\langle \chi\psi_{4}(k,\cdot,t),\psi
_{4^{\prime}}(k,\cdot,t)\right\rangle =a(t)+\mathcal{J}_{1}(t)+\mathcal{J}%
_{2}(t)\,.\label{A_p1}%
\end{equation}
In what follows the asymptotic analysis of these contribution as
$h\rightarrow0$ is developed. Let start with $a(t)$: it can be rephrased as%
\begin{equation}
a(t)=\int\frac{dk}{2\pi h}\,g(k)\,\left\vert \beta(k,t)\right\vert
^{2}\,\left\vert \mu(t)\right\vert ^{2}\,\left\langle \chi
G(t),G(t)\right\rangle _{L^{2}(\mathbb{R})}\,,\label{a_reduc1}%
\end{equation}%
\begin{equation}
\beta(k,t)=C(k,t)-\int_{0}^{t}\dot{C}(k,s)\,e^{-\frac{i}{\varepsilon}\int
_{s}^{t}\left(  E(\sigma)-k^{2}\right)  \,d\sigma}ds\,.\label{Beta}%
\end{equation}
According to the definitions of $\mu$ (see (\ref{Mu_t})) and $\beta$, we have%
\begin{equation}
\partial_{t}\mu(t)=-\frac{\left\langle G(t),\partial_{t}G(t)\right\rangle
}{\left\Vert G(t)\right\Vert _{L^{2}(\mathbb{R})}^{2}}\,\mu(t)+\mathcal{\tilde
{O}}(\varepsilon)\,,\label{der2}%
\end{equation}%
\begin{equation}
\partial_{t}\beta(k,t)=-\frac{i}{\varepsilon}\left(  E(t)-k^{2}\right)
\left(  \beta(k,t)-C(k,t)\right)  \,.\label{der1}%
\end{equation}
Out of exponentially small terms, this leads to the differential relation%
\begin{align}
\medskip\partial_{t}a(t) &  =\left[  -2\operatorname{Re}\frac{\left\langle
G(t),\partial_{t}G(t)\right\rangle }{\left\Vert G(t)\right\Vert _{L^{2}%
(\mathbb{R})}^{2}}+\partial_{t}\ln\left\langle \chi G(t),G(t)\right\rangle
_{L^{2}(\mathbb{R})}\right]  \,a(t)\nonumber\\
&  \medskip-2\operatorname{Re}\frac{i}{\varepsilon}\int\frac{dk}{2\pi
h}\,g(k)\left(  E(t)-k^{2}\right)  \,\left\vert \beta(k,t)\right\vert
^{2}\,\left\vert \mu(t)\right\vert ^{2}\,\left\langle \chi
G(t),G(t)\right\rangle _{L^{2}(\mathbb{R})}\nonumber\\
&  +2\operatorname{Re}\frac{i}{\varepsilon}\int\frac{dk}{2\pi h}\,g(k)\left(
E(t)-k^{2}\right)  \,\left\vert \mu(t)\right\vert ^{2}\,\bar{\beta
}(k,t)C(k,t)\,\left\langle \chi G(t),G(t)\right\rangle _{L^{2}(\mathbb{R})}\,.
\end{align}
Using $E(t)-k^{2}=\left(  E_{R}(t)-k^{2}\right)  -i\Gamma_{t}$, it follows%
\begin{align}
\medskip\partial_{t}a(t) &  =\left[  -2\operatorname{Re}\frac{\left\langle
G(t),\partial_{t}G(t)\right\rangle }{\left\Vert G(t)\right\Vert _{L^{2}%
(\mathbb{R})}^{2}}+\partial_{t}\ln\left\langle \chi G(t),G(t)\right\rangle
_{L^{2}(\mathbb{R})}-2\frac{\Gamma_{t}}{\varepsilon}\right]
\,a(t)+\mathcal{S}^{h}(t)\,,\label{a_reduc.eq}\\[0.4cm]
\mathcal{S}^{h}(t) &  =2\operatorname{Re}\frac{i}{\varepsilon}\int\frac
{dk}{2\pi h}\,g(k)\left(  E(t)-k^{2}\right)  \,\left\vert \mu(t)\right\vert
^{2}\,\bar{\beta}(k,t)C(k,t)\,\left\langle \chi G(t),G(t)\right\rangle
_{L^{2}(\mathbb{R})}\,.\label{a_source_h}%
\end{align}
The derivative at the l.h.s. is explicitly given by%
\begin{equation}
\partial_{t}\ln\left\langle \chi G(t),G(t)\right\rangle _{L^{2}(\mathbb{R}%
)}=2\operatorname{Re}\frac{\left\langle \chi G(t),\partial_{t}%
G(t)\right\rangle }{\left\Vert G(t)\right\Vert _{L^{2}(\mathbb{R})}^{2}}\,,
\end{equation}
so we get%
\begin{equation}
\partial_{t}a(t)=\left[  \frac{2}{\left\vert G(t)\right\vert _{2}^{2}%
}\operatorname{Re}\left\langle \left(  \chi-1\right)  G(t),\partial
_{t}G(t)\right\rangle -2\frac{\Gamma_{t}}{\varepsilon}\right]
\,a(t)+\mathcal{S}^{h}(t)\,.\label{a_reduc.eq1}%
\end{equation}

We next discuss the small-$h$ behaviour of the source term. This can be
further developed as $\mathcal{S}^{h}=\mathcal{S}_{1}^{h}+\mathcal{S}_{2}^{h}%
$\thinspace, with%
\begin{equation}
\mathcal{S}_{1}^{h}(t)=\mathcal{W}(t)\frac{\Gamma_{t}}{\varepsilon}\int
\frac{dk}{2\pi h}\,g(k)\,\left\vert C(k,t)\right\vert ^{2}\,, \label{S_1}%
\end{equation}%
\begin{equation}
\mathcal{S}_{2}^{h}(t)=\mathcal{W}(t)\,\operatorname{Re}\int\frac{dk}{2\pi
h}\,g(k)\,\frac{i}{\varepsilon}\left(  k^{2}-E(t)\right)  \,C(k,t)\,\int
_{0}^{t}\dot{C}^{\ast}(k,s)\,e^{-\frac{i}{\varepsilon}\int_{s}^{t}\left(
k^{2}-E^{\ast}(\sigma)\right)  \,d\sigma}ds\,, \label{S_2}%
\end{equation}
and $\mathcal{W}(t)=2\,\left\vert \mu(t)\right\vert ^{2}\left\langle \chi
G(t),G(t)\right\rangle _{L^{2}(\mathbb{R})}$. Since%
\begin{equation}
\left\vert \mu(t)\right\vert =\frac{\left\Vert G(0)\right\Vert _{L^{2}%
(\mathbb{R})}}{\left\Vert G(t)\right\Vert _{L^{2}(\mathbb{R})}}%
+\mathcal{\tilde{O}}(\varepsilon), \label{Mu_t_mod}%
\end{equation}
the exponentially decreasing character of the Green's functions outside supp
$\chi$ (see Lemma \ref{Lemma_1}) and the relation (\ref{Green_prod}) lead to%
\begin{align}
\mathcal{W}(t)  &  =2\frac{\left\langle \chi G(t),G(t)\right\rangle
_{L^{2}(\mathbb{R})}}{\left\langle G(t),G(t)\right\rangle _{L^{2}(\mathbb{R}%
)}}\left\Vert G(0)\right\Vert _{L^{2}(\mathbb{R})}^{2}+\mathcal{\tilde{O}%
}(\varepsilon)=2\left\Vert G(0)\right\Vert _{L^{2}(\mathbb{R})}^{2}%
+\mathcal{\tilde{O}}\left(  \varepsilon\right) \nonumber\\
&  =\frac{2}{h\alpha_{0}M(E(0),E(0))}+\mathcal{\tilde{O}}\left(
\varepsilon\right)  =\frac{4}{h\left\vert \alpha_{0}\right\vert ^{3}%
}+\mathcal{\tilde{O}}\left(  \varepsilon\right)  \,. \label{factor}%
\end{align}
Thus, $\frac{h\left\vert \alpha_{0}\right\vert ^{3}}{4}\mathcal{S}_{1}^{h}(t)$
expands as (\ref{Int_1}) or (\ref{Int_1.1}), depending on the value of
$d(c,\left\{  a,b\right\}  )$. For the first contribution to $\mathcal{S}^{h}$
we get%
\begin{align}
\medskip\mathcal{S}_{1}^{h}(t)  &  =2\left\vert \frac{\alpha_{t}}{\alpha_{0}%
}\right\vert ^{3}\frac{\Gamma_{t}}{\varepsilon}\,g\left(  \lambda_{t}%
^{\frac{1}{2}}\right)  \left(  1+\mathcal{O}\left(  \left\vert \theta
_{0}\right\vert \right)  \right)  +o(\varepsilon)\,,\quad\qquad\text{for
}d(c,\left\{  a,b\right\}  )=c-a\label{S_1_est}\\[0.4cm]
\mathcal{S}_{1}^{h}(t)  &  =\mathcal{O}\left(  e^{-\frac{2\beta}{h}}\right)
,\qquad\qquad\qquad\qquad\qquad\qquad\qquad\qquad\qquad\quad\text{for
}d(c,\left\{  a,b\right\}  )=b-c \label{S_1_est1}%
\end{align}
with $\beta=\frac{\left\vert \alpha_{t}\right\vert }{h}(c-a-(b-c))$. After
changing the variable: $E=k^{2}$, the second contribution writes as%
\begin{equation}
\mathcal{S}_{2}^{h}(t)=\mathcal{W}(t)\,\operatorname{Re}\int F(E,t)\,dE\,,
\end{equation}%
\begin{equation}
F(E,t)=\frac{1}{2E^{\frac{1}{2}}}\,g(E^{\frac{1}{2}})\,\frac{i}{\varepsilon
}\left(  E-E(t)\right)  \,C(E^{\frac{1}{2}},t)\,\int_{0}^{t}\dot{C}^{\ast
}(\left(  E^{\ast}\right)  ^{\frac{1}{2}},s)\,e^{-\frac{i}{\varepsilon}%
\int_{s}^{t}\left(  E-E^{\ast}(\sigma)\right)  \,d\sigma}ds\,. \label{F_E.t}%
\end{equation}
According to the assumption (h3), $F(\cdot,t)$ extends to an holomorphic
function of $E\in\left\{  z\in\mathbb{C},\ \left\vert z-\lambda_{0}\right\vert
<\frac{h}{d_{0}}\right\}  $, while for $E\in$supp $g(E^{\frac{1}{2}}%
)\diagdown\left\{  \left\vert E-\lambda_{0}\right\vert <\frac{h}{d_{0}%
}\right\}  $, the definitions (\ref{C_kt_exp}), (\ref{C_kt_dot}) and the
exponential bounds (\ref{gen_eigenf^2_exp0}) imply: $\left\vert
F(E,t)\right\vert =\mathcal{\tilde{O}}(\varepsilon)$. In particular, the term:
$\left(  E-E(t)\right)  \,C(E,t)$ is analytic in a complex neighbourhood of
$\lambda_{0}=\lim_{h\rightarrow0}E(0)$, while $\dot{C}^{\ast}(\left(  E^{\ast
}\right)  ^{\frac{1}{2}},s)$ is meromorphic with a double pole at $E=E^{\ast
}(s)$, placed in the upper half plane. Our strategy is to use a complex
integration path formed by the semi-circumference $\mathcal{C}_{\frac{h}{d}%
}(\lambda_{0})$ of center $\lambda_{0}$, radius $\frac{h}{d_{0}}$ in the lower
half-plane: $\operatorname{Im}\mathcal{C}_{\frac{h}{d_{0}}}(\lambda_{0})\leq
0$. Let us consider an holomorphic extension of $F(\cdot,t)$ to the half-disk
whose boundary is determined by: $\left\{  E\in\mathbb{R},\ \left\vert
E-\lambda_{0}\right\vert <\frac{h}{d_{0}}\right\}  \cup\mathcal{C}_{\frac
{h}{d_{0}}}(\lambda_{0})$. Using (\ref{C_kt_exp}), (\ref{C_kt_dot}),
(\ref{gen_eigenf^2_exp2}) and the function $\vartheta\left(  \cdot\right)  $%
\begin{equation}
\vartheta\left(  z\right)  =\left\vert \operatorname{Im}z\right\vert \,,
\label{theta_fun}%
\end{equation}
the restriction of $F(\cdot,t)$ to $\mathcal{C}_{\frac{h}{d_{0}}}(\lambda
_{0})$ is bounded by%
\begin{equation}
\left.  F(\cdot,t)\right\vert _{\mathcal{C}_{\frac{h}{d_{0}}}(\lambda_{0}%
)}\leq\frac{C}{\varepsilon}\,e^{-\frac{\left\vert \alpha_{t}\right\vert }%
{h}(c-a)}\,\int_{0}^{t}e^{-\frac{\vartheta\left(  E\right)  }{\varepsilon
}(t-s)}ds\leq C\,\int_{0}^{t}e^{-\frac{\vartheta\left(  E\right)
}{\varepsilon}(t-s)}ds\,. \label{F_E.t_est1}%
\end{equation}
for a suitable positive $C$. According to (\ref{F_E.t_est1}), the following
estimates hold%
\begin{align*}
\bigskip1.\qquad\left.  F(\cdot,t)\right\vert _{\mathcal{C}_{\frac{h}{d_{0}}%
}(\lambda_{0})}  &  \leq C\frac{\varepsilon}{\vartheta(E)}\left[
1-e^{-\frac{\vartheta\left(  E\right)  }{\varepsilon}t}\right]  \leq
C\frac{\varepsilon}{\vartheta(E)}\,,\\
2.\qquad\left.  F(\cdot,t)\right\vert _{\mathcal{C}_{\frac{h}{d_{0}}}%
(\lambda_{0})}  &  \leq C\,t\,,
\end{align*}
and by interpolation we obtain%
\begin{equation}
\left.  F(\cdot,t)\right\vert _{\mathcal{C}_{\frac{h}{d_{0}}}(\lambda_{0}%
)}\leq C\frac{\varepsilon^{\frac{1}{2}}}{\vartheta^{\frac{1}{2}}(E)}\,.
\label{F_E.t_est3}%
\end{equation}
By computing the residue, $\mathcal{S}_{2}^{h}(t)$ writes as%
\begin{equation}
\mathcal{S}_{2}^{h}(t)=-\mathcal{W}(t)\,\int_{\mathcal{C}_{\frac{h}{d_{0}}%
}(\lambda_{0})}F(E,t)\,dE+\mathcal{\tilde{O}}(\varepsilon)\,. \label{S_2.1}%
\end{equation}
Denoting $z\in\mathcal{C}_{\frac{h}{d_{0}}}(\lambda_{0})$ as: $z=\lambda
_{0}+\frac{h}{d}e^{i\varphi}$, $\omega\in\left(  -\pi,0\right)  $, the
previous inequality implies%
\begin{equation}
\sup_{t}\left\vert \mathcal{S}_{2}^{h}(t)\right\vert \leq C\varepsilon
^{\frac{1}{2}}\int_{\mathcal{C}_{\frac{h}{d_{0}}}(\lambda_{0})}\frac
{\left\vert dE\right\vert }{\vartheta^{\frac{1}{2}}(E)}=C\varepsilon^{\frac
{1}{2}}\frac{h}{d}\int_{-\pi}^{0}\frac{d\omega}{\sin^{\frac{1}{2}}\omega
}=\mathcal{O}(h\,\varepsilon^{\frac{1}{2}})\,. \label{S_2_est}%
\end{equation}

The estimates (\ref{S_1_est})-(\ref{S_1_est1}) and (\ref{S_2_est}), allow to
use $\mathcal{S}^{h}=$ $\mathcal{S}_{1}^{0}+\mathcal{O}\left(  \left\vert
\theta_{0}\right\vert \right)  +\mathcal{O}(h\,\varepsilon^{\frac{1}{2}})$,
with%
\begin{align}
\mathcal{S}_{1}^{0}  &  =2\left\vert \frac{\alpha_{t}}{\alpha_{0}}\right\vert
^{3}\frac{\Gamma_{t}}{\varepsilon}\,g\left(  \lambda_{t}^{\frac{1}{2}}\right)
,\qquad\text{for }d(c,\left\{  a,b\right\}  )=c-a\,,\label{S_1_lim}\\
\mathcal{S}_{1}^{0}  &  =\mathcal{O}\left(  e^{-\frac{\beta}{h}}\right)
\,,\qquad\qquad\quad\ \ \text{for }d(c,\left\{  a,b\right\}  )=b-c\,,
\label{S_1_lim1}%
\end{align}
with: $\beta=\frac{\left\vert \alpha_{t}\right\vert }{h}(c-a-(b-c))$. Owing to
the estimates in Lemma \ref{Lemma_1}, the term $\frac{2}{\left\vert
G(t)\right\vert _{2}^{2}}\operatorname{Re}\left\langle \left(  \chi-1\right)
G(t),\partial_{t}G(t)\right\rangle $\,, is bounded by%
\begin{equation}
\frac{1}{\left\vert G(t)\right\vert _{2}^{2}}\left\vert \left\langle \left(
\chi-1\right)  G(t),\partial_{t}G(t)\right\rangle _{L^{2}(\mathbb{R}%
)}\right\vert \leq C\left\vert \left(  \chi-1\right)  G(t)\right\vert
_{L^{2}(\mathbb{R})}\left\vert \partial_{t}G(t)\right\vert _{L^{2}%
(\mathbb{R})}=\mathcal{\tilde{O}}\left(  \inf_{\text{supp }(1-\chi)}%
e^{-\frac{\left\vert \alpha_{0}\right\vert }{2h}\left\vert \cdot-c\right\vert
}\right)  \,. \label{a_potential_est}%
\end{equation}

When the interaction point '$c$' is on the left side of the barrier's support
and the condition $d(c,\left\{  a,b\right\}  )=c-a$ is fulfilled, the limit
condition (\ref{S_1_lim}) and the estimates (\ref{S_2_est}),
(\ref{a_potential_est}) allow to write (\ref{a_reduc.eq1}) as follows%
\begin{equation}
\partial_{t}a(t)=\left(  -2\frac{\Gamma_{t}}{\varepsilon}\right)  \,\left(
a(t)-\left\vert \frac{\alpha_{t}}{\alpha_{0}}\right\vert ^{3}g\left(
\lambda_{t}^{\frac{1}{2}}\right)  \right)  +\mathcal{O}\left(  \left\vert
\theta_{0}\right\vert \right)  +\mathcal{\tilde{O}}\left(  e^{-\frac
{\tau_{\chi}}{h}}\right)  \,, \label{a_reduc.eq2}%
\end{equation}
where $\tau_{\chi}>0$ is defined according to the remainders in (\ref{S_1_est}%
), (\ref{S_2_est}) and (\ref{a_potential_est}). The initial datum for this
equation is deduced by evaluating (\ref{a_reduc1}) at $t=0$. With the above
expansions, we obtain%
\begin{equation}
a(0)=g\left(  \lambda_{0}^{\frac{1}{2}}\right)  \left(  1+\mathcal{O}\left(
\left\vert \theta_{0}\right\vert \right)  \right)  +o\left(  \varepsilon
\right)  \,,\qquad\text{for }d(c,\left\{  a,b\right\}  )=c-a\,. \label{a_0}%
\end{equation}
When $d(c,\left\{  a,b\right\}  )=c-a$\,, the solution $a(t)$ is%
\begin{equation}
\left\{
\begin{array}
[c]{l}%
\medskip a(t)=a(0)e^{-2\int_{0}^{t}\frac{\Gamma_{s}}{\varepsilon}ds}+\int
_{0}^{t}e^{-2\int_{s}^{t}\frac{\Gamma_{\sigma}}{\varepsilon}d\sigma
}\mathcal{S}_{1}^{0}(s)\,ds+\mathcal{O}\left(  \left\vert \theta
_{0}\right\vert \right)  +\mathcal{\tilde{O}}\left(  e^{-\frac{\tau_{\chi}}%
{h}}\right)  \,,\\
a(0)=\,g\left(  \lambda_{0}^{\frac{1}{2}}\right)  \,;\qquad\mathcal{S}_{1}%
^{0}(t)=2\left\vert \frac{\alpha_{t}}{\alpha_{0}}\right\vert ^{3}\frac
{\Gamma_{t}}{\varepsilon}\,g\left(  \lambda_{t}^{\frac{1}{2}}\right)  \,.
\end{array}
\right.  \label{a_reduc.eq3}%
\end{equation}
In the other case, when $d(c,\left\{  a,b\right\}  )=b-c$, the initial value
of $a(t)$ is: $a(0)=\mathcal{O}\left(  e^{-\frac{\beta}{h}}\right)  $ which
coincides with the size of the source term given in (\ref{S_1_lim1}). This
leads to%
\begin{equation}
a(t)=\mathcal{O}\left(  e^{-\frac{\beta}{h}}\right)  ,\qquad\text{for
}d(c,\left\{  a,b\right\}  )=b-c. \label{a_right}%
\end{equation}
with $\beta=\frac{\left\vert \alpha_{t}\right\vert }{h}(c-a-(b-c))$.

To complete the proof of the second point of Theorem~\ref{Th_1}, we need the
following Lemma.

\begin{lemma}
\label{Lemma_outdiag}In the assumptions (h1)-(h4), the relations%
\begin{equation}
\int\frac{dk}{2\pi h}\,g(k)\,\left\langle \chi\,\psi_{j}(k,\cdot
,t),\psi_{j^{\prime}}(k,\cdot,t)\right\rangle _{L^{2}(\mathbb{R}%
)}=\mathcal{\tilde{O}}(\varepsilon^{\frac{1}{2}}),
\end{equation}
hold with: $j,j^{\prime}=1,2,3,4$ and $j\neq j^{\prime}$\,.
\end{lemma}

\begin{preuve}
Let consider the contributions $J_{i=1,2}(t)$ to (\ref{A_p}). The first term
explicitly writes as%
\begin{equation}
\mathcal{J}_{1}(t)=\left\vert 1-\mu(t)\right\vert ^{2}\left\langle \chi
G(t),G(t)\right\rangle _{L^{2}(\mathbb{R})}\int\frac{dk}{2\pi h}%
\,g(k)\,\left\vert C(k,t)\right\vert ^{2}\,.
\end{equation}
Then, the estimate (\ref{Green_uplow}), and the relations (\ref{Int_1}),
(\ref{Mu_t_mod}) yield: $\mathcal{J}_{1}(t)=\mathcal{\tilde{O}}(\varepsilon
^{0})$, holding for any choice of $\chi,g$ fulfilling the assumptions. For the
second terms, let take $\tilde{g}$, $\tilde{\chi}$ a couple of positive
functions fulfilling (h3), and such that: $\left\vert g\right\vert <\tilde{g}%
$, $\left\vert \chi\right\vert <\tilde{\chi}$. With this conditions, a
straightforward application of the Cauchy-Schwartz inequality gives%
\begin{align*}
\frac{\left\vert \mathcal{J}_{2}(t)\right\vert }{2} &  \leq\left\vert
\int\frac{dk}{2\pi h}\,g(k)\,\left\langle \chi\varphi_{1}(k,\cdot
,t),\varphi_{2}(k,\cdot,t)\right\rangle \right\vert \leq\int\int\frac
{dk\,dx}{2\pi h}\,\tilde{g}(k)\tilde{\chi}(x)\,\left\vert \varphi
_{1}(k,x,t),\varphi_{2}(k,x,t)\right\vert \\
&  \leq\left(  \int\frac{dk}{2\pi h}\,\tilde{g}(k)\,\left\langle \tilde{\chi
}\varphi_{1}(k,\cdot,t),\varphi_{1}(k,\cdot,t)\right\rangle \right)
^{\frac{1}{2}}\left(  \int\frac{dk}{2\pi h}\,\tilde{g}(k)\,\left\langle
\tilde{\chi}\varphi_{2}(k,\cdot,t),\varphi_{2}(k,\cdot,t)\right\rangle
\right)  ^{\frac{1}{2}}\\
&  =\tilde{a}^{\frac{1}{2}}(t)\mathcal{\tilde{J}}_{1}^{\frac{1}{2}}(t)\,,
\end{align*}
with $\tilde{a}$ and $\mathcal{\tilde{J}}_{1}$ denoting the principal
contribution and the first remainder arising from the auxiliary data
$\tilde{g}$, $\tilde{\chi}$. Since $\tilde{a}(t)=\mathcal{O}(1)$ (as it
follows from (\ref{a_reduc.eq3})) and $\mathcal{\tilde{J}}_{1}%
(t)=\mathcal{\tilde{O}}(\varepsilon^{0})$, we obtain: $\mathcal{J}%
_{2}(t)=\mathcal{\tilde{O}}(\varepsilon^{0})$. This leads to%
\begin{equation}
\int\frac{dk}{2\pi h}\,g(k)\,\left\langle \chi\,\psi_{4}(k,\cdot,t),\psi
_{4}(k,\cdot,t)\right\rangle _{L^{2}(\mathbb{R})}=\mathcal{\tilde{O}%
}(\varepsilon^{0})\label{A_p_bound}%
\end{equation}
while the results of Lemma \ref{Lemma_diag}, gives
\begin{equation}
\int\frac{dk}{2\pi h}\,g(k)\,\left\langle \chi\,\psi_{j}(k,\cdot,t),\psi
_{j}(k,\cdot,t)\right\rangle _{L^{2}(\mathbb{R})}=\mathcal{\tilde{O}%
}(\varepsilon),\qquad\text{with}\;j=1,2,3\,.\label{A_p_bound1}%
\end{equation}
Once more, we remark that these estimates hold for all choice of $g,\chi$
fulfilling the conditions (h3). Let $g_{m}$ and $\chi_{m}$ be positive
defined, verifying the required hypothesis and such that:\ $g_{m}>\left\vert
g\right\vert $ and $\chi_{m}>\left\vert \chi\right\vert $. For $j\neq
j^{\prime}$, the Cauchy-Schwarz inequality implies%
\begin{multline*}
\left\vert \int\frac{dk}{2\pi h}\,g(k)\,\left\langle \chi\,\psi_{j}%
(k,\cdot,t),\psi_{j^{\prime}}(k,\cdot,t)\right\rangle _{L^{2}(\mathbb{R}%
)}\right\vert \leq\int\int\frac{dk\,dx}{2\pi h}\,g_{m}(k)\,\chi_{m}%
(x)\,\left\vert \psi_{j}(k,x,t)\,\psi_{j^{\prime}}(k,x,t)\right\vert \\
\leq\left(  \int\frac{dk}{2\pi h}\,g_{m}(k)\,\left\langle \chi_{m}\,\psi
_{j}(k,\cdot,t),\psi_{j}(k,\cdot,t)\right\rangle _{L^{2}(\mathbb{R})}\right)
^{\frac{1}{2}}\left(  \int\frac{dk}{2\pi h}\,g_{m}(k)\,\left\langle \chi
_{m}\,\psi_{j^{\prime}}(k,\cdot,t),\psi_{j^{\prime}}(k,\cdot,t)\right\rangle
_{L^{2}(\mathbb{R})}\right)  ^{\frac{1}{2}}\\
\leq\mathcal{\tilde{O}}(\varepsilon^{\frac{1}{2}})\,.
\end{multline*}

\end{preuve}

\subsection{\label{Sec_proof}Remainder terms and proof of Theorem~\ref{Th_1}}

We next consider the terms $\mathcal{J}_{1}(t)$ and $\mathcal{J}_{2}(t)$ in
(\ref{A_p1}) in the limit $h\rightarrow0$. To this aim, an asymptotic formula
for the difference: $1-\mu(t)$ is needed.

\begin{lemma}
With the assumptions (h1)-(h4), the function $\mu(t)$, defined in
(\ref{Mu_t}), is such that%
\begin{equation}
\mu(t)=\left\vert \frac{\alpha_{t}}{\alpha_{0}}\right\vert ^{\frac{3}{2}%
}\left(  1+\mathcal{\tilde{O}}\left(  \varepsilon\right)  \right)
=1+\mathcal{O}(h)\,.\label{Mu_t_exp}%
\end{equation}

\end{lemma}

\begin{preuve}
From (\ref{Mu_t}) and (\ref{Mu_t_mod}), our function writes as%
\begin{equation}
\mu(t)=\frac{\left\Vert G(0)\right\Vert _{L^{2}(\mathbb{R})}}{\left\Vert
G(t)\right\Vert _{L^{2}(\mathbb{R})}}e^{-i\int_{0}^{t}\frac{\operatorname{Im}%
\left\langle G(s),\partial_{s}G(s)\right\rangle }{\left\vert G(s)\right\vert
_{2}^{2}}ds}\,+\mathcal{\tilde{O}}(\varepsilon). \label{Mu1}%
\end{equation}
As $h\rightarrow0$, an approximation of $\operatorname{Im}\left\langle
G(s),\partial_{s}G(s)\right\rangle $ is computable starting from the relation
(\ref{Green_prod_diff}) taken with: $E=E_{res}^{h}=E(s)$ and $\alpha
=\alpha_{s}$; this gives%
\[
\operatorname{Im}\left\langle G(s),\partial_{s}G(s)\right\rangle =-\frac
{1}{h\alpha_{s}}\operatorname{Im}\frac{\dot{E}(s)\partial_{2}M(E(s),E(s))}%
{M^{2}(E(s),E(s))}+\mathcal{\tilde{O}}\left(  \varepsilon\right)  \,,
\]
where $\partial_{2}$ denotes the derivative w.r.t the second variable. A
relation for $\dot{E}(t)$ follows by taking the time derivative of
(\ref{eigenvalue_eq1}),%
\begin{equation}
\dot{E}(t)=\frac{\dot{\alpha}_{t}}{\alpha_{t}}\frac{G^{E(t)}(c,c)}{\left.
\partial_{E}G^{E}(c,c)\right\vert _{E(t)}}\,. \label{E_res_dot}%
\end{equation}
The r.h.s. of (\ref{E_res_dot}) is further developed by using (\ref{M_E_exp});
this leads to: $\dot{E}(t)=\frac{\dot{\alpha}_{t}\left\vert \alpha
_{t}\right\vert }{2}+$ $\mathcal{O}\left(  e^{-\frac{\left\vert \alpha
_{t}\right\vert }{h}d(c,\left\{  a,b\right\}  )}\right)  $. Thus, $\dot{E}(t)$
is real, out of exponentially small terms, and the size of $\operatorname{Im}%
\left\langle G(s),\partial_{s}G(s)\right\rangle $ is determined by the
imaginary part of $M^{-2}(E(s),E(s))\,\partial_{2}M(E(s),E(s))$. According to
(\ref{M_E}), this quantity expresses as%
\[
\frac{\partial_{2}M(E(s),E(s))}{M^{2}(E(s),E(s))}=\frac{\frac{1}%
{2}+\mathcal{\tilde{O}}\left(  e^{-\frac{\left\vert \alpha_{s}\right\vert }%
{h}d(c,\left\{  a,b\right\}  )}\right)  }{\left\vert -\frac{\alpha_{s}^{2}}%
{2}+\mathcal{\tilde{O}}\left(  e^{-\frac{\left\vert \alpha_{s}\right\vert }%
{h}d(c,\left\{  a,b\right\}  )}\right)  \right\vert ^{2}}=-\frac{2}{\alpha
_{s}^{4}}+\mathcal{\tilde{O}}\left(  \varepsilon\right)  \,.
\]
We finally get: $\operatorname{Im}\left\langle G(s),\partial_{s}%
G(s)\right\rangle =\mathcal{\tilde{O}}\left(  e^{-\frac{\left\vert \alpha
_{s}\right\vert }{h}d(c,\left\{  a,b\right\}  )}\right)  $. It follows that%
\begin{equation}
\mu(t)=\frac{\left\Vert G(0)\right\Vert _{L^{2}(\mathbb{R})}}{\left\Vert
G(t)\right\Vert _{L^{2}(\mathbb{R})}}\left(  1+\mathcal{\tilde{O}}\left(
\varepsilon\right)  \right)  \,. \label{Mu_t_exp1}%
\end{equation}
Since the Green's functions norms can be expressed in terms of $\left(
h\alpha_{t}M(E(t),E(t))\right)  ^{-1}$ (we refer to (\ref{Green_prod})), the
above ratio further expands as%
\begin{equation}
\frac{\left\Vert G(0)\right\Vert _{L^{2}(\mathbb{R})}}{\left\Vert
G(t)\right\Vert _{L^{2}(\mathbb{R})}}=\left\vert \frac{\alpha_{t}}{\alpha_{0}%
}\right\vert ^{\frac{3}{2}}+\mathcal{\tilde{O}}\left(  \varepsilon\right)  \,.
\end{equation}
This result, together with the assumption (\ref{alpha_var}), leads to
(\ref{Mu_t_exp}).
\end{preuve}

The integral $\mathcal{J}_{1}(t)$ has the form%
\begin{equation}
\mathcal{J}_{1}(t)=\left\vert 1-\mu(t)\right\vert ^{2}\left\langle \chi
G(t),G(t)\right\rangle _{L^{2}(\mathbb{R})}\int\frac{dk}{2\pi h}%
\,g(k)\,\left\vert C(k,t)\right\vert ^{2}\,.\label{J_1}%
\end{equation}
According to (\ref{Int_1})-(\ref{Int_1.1}), (\ref{Green_prod}%
),(\ref{Mu_t_exp1}), and using the exponential estimates for $G(t)$ outside
supp $\chi$, this can be rephrased as%
\begin{equation}
\mathcal{J}_{1}(t)=\left\vert 1-\left\vert \frac{\alpha_{t}}{\alpha_{0}%
}\right\vert ^{\frac{3}{2}}\right\vert ^{2}\,g\left(  \lambda_{t}^{\frac{1}%
{2}}\right)  \left(  1+\mathcal{O}\left(  \left\vert \theta_{0}\right\vert
\right)  \right)  +\mathcal{\tilde{O}}\left(  e^{-\frac{\tau}{h}}\right)
\,,\label{J_1_exp}%
\end{equation}
for a suitable $\tau>0$ and $d(c,\left\{  a,b\right\}  )=c-a$, otherwhise we
have: $\mathcal{J}_{1}=\mathcal{O}\left(  e^{-\frac{\beta}{h}}\right)  $. The
second remainder is a crossing term (see definition: \ref{J2}); in Lemma
\ref{Lemma_outdiag} it has been shown that: $\mathcal{J}_{2}=\mathcal{O}%
\left(  \tilde{a}^{\frac{1}{2}}\,\mathcal{\tilde{J}}_{1}^{\frac{1}{2}}\right)
$ where the variables $\tilde{a}$ and $\mathcal{\tilde{J}}_{1}$ are the
principal contribution and the first remainder associated with a suitable
couple of auxiliary data $\tilde{g}$, $\tilde{\chi}$. If we assume
$d(c,\left\{  a,b\right\}  )=b-c$, we have: $\mathcal{J}_{2}\sim\tilde{a}%
\cdot\mathcal{\tilde{J}}_{1}=\mathcal{O}\left(  e^{-\frac{\beta}{h}}\right)
$. When $d(c,\left\{  a,b\right\}  )=c-a$, this term is explicitely given by%
\begin{equation}
\mathcal{J}_{2}(t)=2\operatorname{Re}\mu(t)\left(  1-\mu^{\ast}(t)\right)
\,\left\langle \chi G(t),G(t)\right\rangle _{L^{2}(\mathbb{R})}\,\,\int
\frac{dk}{2\pi h}\,g(k)\,\beta(k,t)C^{\ast}(k,t)\,.\label{J_2}%
\end{equation}
After an integration by part, we get%
\begin{equation}
\mathcal{J}_{2}(t)=2\operatorname{Re}\mu(t)\left(  1-\mu^{\ast}(t)\right)
\,\left\langle \chi G(t),G(t)\right\rangle _{L^{2}(\mathbb{R})}\,\left[
I+II\right]  \,,\label{J_2_1}%
\end{equation}%
\begin{equation}
I=\int\frac{dk}{2\pi h}\,g(k)\,C(k,0)C^{\ast}(k,t)e^{-\frac{i}{\varepsilon
}\int_{0}^{t}\left(  E(\sigma)-k^{2}\right)  \,d\sigma}\,,\label{I}%
\end{equation}%
\begin{equation}
II=\frac{i}{\varepsilon}\int\frac{dk}{2\pi h}\,g(k)\,\int_{0}^{t}%
C(k,s)C^{\ast}(k,t)\left(  E(s)-k^{2}\right)  e^{-\frac{i}{\varepsilon}%
\int_{s}^{t}\left(  E(\sigma)-k^{2}\right)  \,d\sigma}ds\,.\label{II}%
\end{equation}
The small-$h$ behaviour of $I$ and $II$ is investigated using a
path-deformation argument and following the same line as in (\ref{S_2}). As
before, $\mathcal{C}_{\frac{h}{d_{0}}}^{+}(\lambda_{0})$ denotes the
semicircle of center $\lambda_{0}$, radius $\frac{h}{d_{0}}$, but now we fix
$\operatorname{Im}\mathcal{C}_{\frac{h}{d_{0}}}^{+}(\lambda_{0}\mathcal{)}>0$.
Replacing $C(k,0)C^{\ast}(k,t)$ with $C(k,0)C^{\ast}(k^{\ast},t)$, we define a
meromorphic function in a neighbourhood of $\lambda_{0}$ with simple poles at
$k^{2}=E(0),E^{\ast}(t)$. Thus, the first integral is%
\begin{equation}
I=-\int_{\mathcal{C}_{\frac{h}{d_{0}}}^{+}(\lambda_{0})}\frac{dE}{4h\pi
E^{\frac{1}{2}}}\,g\left(  E^{\frac{1}{2}}\right)  \mathcal{K}%
(E,0,t)\,e^{-\frac{i}{\varepsilon}\int_{0}^{t}\left(  E(\sigma)-E\right)
\,d\sigma}\,dE+2\pi i\emph{\,Res}_{1}(E^{\ast}(t))+\mathcal{\tilde{O}%
}(\varepsilon)\,,
\end{equation}
where $\emph{Res}_{1}(E^{\ast}(t))$ is the residue at $E^{\ast}(t)$, while
$\mathcal{K}(E,s,t)$ denotes%
\[
\mathcal{K}(E,s,t)=C\left(  E^{\frac{1}{2}},s\right)  C^{\ast}\left(  \left(
E^{\ast}\right)  ^{\frac{1}{2}},t\right)  \,.
\]
Since $\operatorname{Im}\left(  E(\sigma)-k^{2}\right)  <0$ and $\mathcal{K}%
(E,s,t)=\mathcal{\tilde{O}}(\varepsilon)$ for $E\in\mathcal{C}_{\frac{h}%
{d_{0}}}^{+}(\lambda_{0})$ (according to (\ref{gen_eigenf^2_exp2})), we have%
\begin{equation}
I=2\pi i\emph{\,Res}_{1}(E^{\ast}(t))+\mathcal{\tilde{O}}(\varepsilon
)\,.\label{J_2_1.1}%
\end{equation}
Computing the residue when $h\rightarrow0$, $E(t)$ can be replaced with its
limit value $\lambda_{t}$, excepting those parts of the function where the
difference $E^{\ast}(t)-E(0)$ appears. In this case we use: $E(t)=\lambda
_{t}-i\Gamma_{t}$. Out of exponentially small terms, the result is%
\begin{equation}
\emph{Res}_{1}(E^{\ast}(t))=\frac{h\alpha_{0}\alpha_{t}}{4\pi\lambda
_{t}^{\frac{1}{2}}}\,g\left(  \lambda_{t}^{\frac{1}{2}}\right)  M(\lambda
_{t},\lambda_{0})M^{\ast}(\lambda_{t},\lambda_{t})\frac{\left\vert \tilde
{\psi}_{-}(\lambda_{t}^{\frac{1}{2}},c)\right\vert ^{2}}{E^{\ast}%
(t)-E(0)}\,e^{-\frac{i}{\varepsilon}\int_{0}^{t}\left(  E(\sigma)-E^{\ast
}(t)\right)  \,d\sigma}\,.
\end{equation}
Using (\ref{M_E}) and (\ref{gen_eigenf^2_exp1}), it follows%
\begin{equation}
\,2\pi i\emph{Res}_{1}(E^{\ast}(t))=i\left\vert \alpha_{0}\right\vert \left(
\alpha_{t}^{2}+\alpha_{0}\alpha_{t}\right)  \,g\left(  \lambda_{t}^{\frac
{1}{2}}\right)  \frac{\,\,e^{-\frac{i}{\varepsilon}\int_{0}^{t}\left(
E(\sigma)-E^{\ast}(t)\right)  \,d\sigma}}{E^{\ast}(t)-E(0)}\frac{\Gamma_{t}%
}{2}\,\left(  h+\mathcal{O}\left(  \left\vert h\,\theta_{0}\right\vert
\right)  \right)  \,.\label{J_2_res1}%
\end{equation}

Adopting the same notation, the second contribution writes as%
\begin{multline}
II=-\frac{i}{\varepsilon}\int_{\mathcal{C}_{\frac{h}{d_{0}}}^{+}(\lambda_{0}%
)}\frac{dE}{4\pi hE^{\frac{1}{2}}}\,g(E^{\frac{1}{2}})\,\int_{0}%
^{t}\mathcal{K}(E,s,t)\left(  E(s)-E\right)  \,e^{-\frac{i}{\varepsilon}%
\int_{s}^{t}\left(  E(\sigma)+E\right)  \,d\sigma}ds\\
+2\pi i\,\emph{Res}_{2}(E^{\ast}(t))+\mathcal{\tilde{O}}(\varepsilon)
\end{multline}
Proceeding as the previous Section (see the estimate of $\mathcal{S}_{2}%
^{h}(t)$), the integral over $\mathcal{C}_{\frac{h}{d_{0}}}^{+}(\lambda_{0})$
is bounded as $\mathcal{O}(\varepsilon^{\frac{1}{2}})$, while the residue in
$E^{\ast}(t)$ is given, out of exponentially small terms, by%
\[
\,\emph{Res}_{2}(E^{\ast}(t))=-i\frac{h}{4\pi}\,g(\lambda_{t}^{\frac{1}{2}%
})\frac{\Gamma_{t}}{\varepsilon}\left(  1+\mathcal{O}\left(  \left\vert
\theta_{0}\right\vert \right)  \right)  \,\int_{0}^{t}f(s,t)\,e^{-\frac
{i}{\varepsilon}\varphi(s,t)}ds\,,
\]%
\[
f(s,t)=\left(  \alpha_{s}\alpha_{t}^{2}+\alpha_{s}^{2}\alpha_{t}\right)
\,e^{-\frac{1}{\varepsilon}\int_{s}^{t}\left(  \Gamma_{\sigma}+\Gamma
_{t}\right)  \,d\sigma}\,;\qquad\varphi(s,t)=\int_{s}^{t}\left(
\lambda_{\sigma}-\lambda_{t}\right)  \,d\sigma\,.
\]
If $d(c,\left\{  a,b\right\}  )=c-a$, the factor $\frac{\Gamma_{t}%
}{\varepsilon}$ is $\mathcal{O}(1)$, and the size of $II$ is determined by the
oscillatory integral. To this concern, we notice that: $\partial_{s}%
\varphi(s,t)=\lambda_{t}-\lambda_{s}$; according to the definition of
$\lambda_{t}$, the stationary points of $\varphi(s,t)$ are defined by the
equation%
\begin{equation}
\alpha_{s}-\alpha_{t}=0\,.\label{stat.point_condition}%
\end{equation}
It follows from (h2) that the set of the '$s$' fulfilling the condition
(\ref{stat.point_condition}) does not have accumulation points in $\left[
0,t\right]  $, forming a subset of finite cardinality. It means that
$s\rightarrow\varphi(s,t)$ have a finite number of stationary points $\left\{
s_{j}(t)\right\}  _{j=1}^{N(t)}\subset$ $\left[  0,t\right]  $, depending on
$t$. Since $s\rightarrow f(s,t)$ is a regular function (with $\alpha
_{s},\Gamma_{s}\in C^{\infty}$) the stationary phase method applies with:
$\left\vert \partial_{s}^{j+1}\varphi(s,t)\right\vert =\left\vert \partial
_{s}^{j}E_{R}(s)\right\vert \gtrsim\left\vert \partial_{t}^{j}\alpha
(t)\right\vert >0$ for some $j\in\left\{  1,...k\right\}  $. This yields:
$\int_{0}^{t}f(s,t)\,e^{-\frac{i}{\varepsilon}\varphi(s,t)}ds=\mathcal{O}%
\left(  \varepsilon^{\frac{1}{j+1}}\right)  $ and%
\begin{equation}
\,\emph{Res}_{2}(E^{\ast}(t))\,=\mathcal{O}\left(  \varepsilon^{\frac{1}{j+1}%
}\right)  ,\label{J_2_res2}%
\end{equation}
and uniformly w.r.t. the time. According to the definition (\ref{J_2_1}) and
the expasions (\ref{J_2_1.1}), (\ref{J_2_res1}), (\ref{J_2_res2}), we get%
\begin{multline*}
\mathcal{J}_{2}(t)=2\operatorname{Re}\left[  i\mu(t)\left(  1-\mu^{\ast
}(t)\right)  \,\left\langle \chi G(t),G(t)\right\rangle _{L^{2}(\mathbb{R}%
)}g\left(  \lambda_{t}^{\frac{1}{2}}\right)  \frac{\Gamma_{t}}{2}\right.
\times\\
\times\left.  \left\vert \alpha_{0}\right\vert \left(  \alpha_{t}^{2}%
+\alpha_{0}\alpha_{t}\right)  \,\frac{\,\,e^{-\frac{i}{\varepsilon}\int
_{0}^{t}\left(  E(\sigma)-E^{\ast}(t)\right)  \,d\sigma}}{E^{\ast}%
(t)-E(0)}\,\left(  h+\mathcal{O}\left(  \left\vert h\,\theta_{0}\right\vert
\right)  \right)  \right]
\end{multline*}
Expanding $\,\left\langle \chi G(t),G(t)\right\rangle _{L^{2}(\mathbb{R})}$
and $\mu(t)$ with (\ref{Green_prod}), (\ref{Mu_t_exp}) and using
$E(t)=\lambda_{t}-i\Gamma_{t}+\mathcal{\tilde{O}}\left(  \varepsilon\right)
$, leads to%
\begin{equation}
\mathcal{J}_{2}(t)=\operatorname{Re}2i\left(  1-\left\vert \frac{\alpha_{t}%
}{\alpha_{0}}\right\vert ^{\frac{3}{2}}\right)  \,\frac{\Gamma_{t}%
}{\varepsilon}\,g\left(  \lambda_{t}^{\frac{1}{2}}\right)  \frac
{\mathcal{T}(t)}{\frac{\lambda_{t}-\lambda_{0}}{\varepsilon}-i\frac{\left(
\Gamma_{t}+\Gamma_{0}\right)  }{\varepsilon}}\,,\label{J_2_exp}%
\end{equation}%
\begin{equation}
\mathcal{T}(t)=\frac{\left\vert \alpha_{0}\right\vert \alpha_{t}^{2}%
+\alpha_{0}^{2}\left\vert \alpha_{t}\right\vert }{\left(  \alpha_{0}\alpha
_{t}\right)  ^{\frac{3}{2}}}\,e^{-\frac{1}{\varepsilon}\int_{0}^{t}\left(
\Gamma_{\sigma}+\Gamma_{t}\right)  \,d\sigma}e^{-\frac{i}{\varepsilon}\int
_{0}^{t}\left(  \lambda_{\sigma}-\lambda_{t}\right)  \,d\sigma}%
\,.\label{J_2_exp1}%
\end{equation}
When $d(c,\left\{  a,b\right\}  )=c-a$, $\Gamma_{t}$ is $\mathcal{O}%
(\varepsilon)$ and the small-$h$ behaviour of this quantity is determined by
the ratio: $\frac{\lambda_{t}-\lambda_{0}}{\varepsilon}$. In particular, for
$\lambda_{t}\neq\lambda_{0}$, one has: $\emph{Res}_{1}(E^{\ast}%
(t))=\mathcal{O}(\varepsilon)$. However, if: $E_{R}(0)-E_{R}(t)\sim
\mathcal{O}(\varepsilon)$, a boundary layer contribution is expected.\bigskip

\begin{preuve}
[Proof of Theorem~\ref{Th_1}]$i)$ This first point is a rewriting of the
result of Proposition \ref{Prop_1}.\newline$ii)$ The second point comes from
the decomposition (\ref{A_decomp}) and the results of Lemmas \ref{Lemma_diag}
and \ref{Lemma_outdiag}. The reduced equation for the main contribution $a(t)$
is obtained in (\ref{a_reduc.eq2}) for $d(c,\left\{  a,b\right\}  )=c-a$,
while this variable is exponentially small, according to the estimate
(\ref{a_right}), when $d(c,\left\{  a,b\right\}  )=b-c$.\newline$iii)$ Once
the small-$h$ behaviour of the factors $\mu(t)$, $\left(  1-\mu^{\ast
}(t)\right)  $ and $\,\left\langle \chi G(t),G(t)\right\rangle _{L^{2}%
(\mathbb{R})}$ is taken into account, the last point is a consequence of
(\ref{J_1_exp}), (\ref{J_2_1}), (\ref{J_2_1.1}) and (\ref{J_2_exp}%
)-(\ref{J_2_exp1}).
\end{preuve}


\bigskip

\begin{thebibliography}{99}                                                                                               %


\bibitem {Albeverio}S.Albeverio, F.Gesztesy, R.H\"{o}gh-Krohn and H.Holden.
\emph{Solvable Models in Quantum Mechanics 2nd ed. with an appendix by P.
Exner}. AMS, Providence R.I, 2005.

\bibitem {AgCo}J. Aguilar, J.M. Combes. A class of analytic perturbations for
one-body Schr\"{o}dinger Hamiltonians. Comm. Math. Phys., \textbf{22},
269--279, 1971.

\bibitem {AEGS}J.E. Avron, A. Elgart, G.M. Graf, L. Sadun. Transport and
dissipation in quantum pumps. \emph{J. Stat. Phys., }\textbf{116}(1-4),
425-473, 2004.

\bibitem {AEGSS}J.E. Avron, A. Elgart, G.M. Graf, L. Sadun, K. Schnee.
Adiabatic charge pumping in open quantum systems. \emph{Comm. Pure Appl.
Math.,} \textbf{57}(4), 528--561, 2004.

\bibitem {BaCo}E. Balslev, J.M. Combes. Spectral properties of many-body
Schr\"{o}dinger operators with dilatation-analytic interactions. Comm. Math.
Phys., \textbf{22}, 280--294, 1971.

\bibitem {BNP1}V. Bonnaillie-No\"{e}l, F. Nier, Y. Patel. Far from equlibrium
steady states of 1D-Schr\"{o}dinger-Poisson systems with quantum wells I. Ann.
I.H.P. An. Non Lin\'{e}aire, \textbf{25}, 937-968, 2008.

\bibitem {BNP2}V. Bonnaillie-No\"{e}l, F. Nier, Y. Patel. Far from equlibrium
steady states of 1D-Schr\"{o}dinger-Poisson systems with quantum wells II. J.
Math. Soc. of Japan., \textbf{61}, 65-106, 2009.

\bibitem {BNP3}V. Bonnaillie-No\"{e}l, F. Nier, Y. Patel. Computing the steady
states for an asymptotic model of quantum transport in resonant
heterostructures. Journal of Computational Physics, \textbf{219}(2), 644-670, 2006.

\bibitem {CDNP}H.D. Cornean, P. Duclos, G. Nenciu, R. Purice. Adiabatically
switched-on electrical bias and the Landauer-B\"{u}ttiker formula. \emph{J.
Math. Phys.,} \textbf{49}(10), 2008.

\bibitem {FMN1}A. Faraj, A.Mantile, F.Nier. Double scale analysis of a
Schr\"{o}dinger-Poisson system with quantum wells and macroscopic
nonlinearities in dimension 2 and 3. Asymptot. Anal. \textbf{62} no. 3-4,
163--205, 2009.

\bibitem {FMN2}A. Faraj, A.Mantile, F.Nier. Adiabatic evolution of 1D shape
resonances: an artificial interface conditions approach. Preprint arXiv:1001.3665.

\bibitem {Hel}B. Helffer. \emph{Semiclassical analysis for the Schr\"{o}dinger
operator and applications}, volume 1336 of \emph{Lecture Notes in Mathematics.
}Springer-Verlag, Berlin, 1988.

\bibitem {HeSj1}B. Helffer, J. Sj\"{o}strand. \emph{R\'{e}sonances en limite
semi-classique. }Number 24-25 in M\'{e}m. Soc. Mat. France (N.S.), 1986.

\bibitem {JoPrSj}G. Jona-Lasionio, C. Presilla, J. Sj\"{o}strand. On the
Schr\"{o}dinger equation with concentrated non linearities. \emph{Ann.
Physics, }\textbf{240}(1), 1-21,1995.

\bibitem {Joye1}A. Joye. General adiabatic evolution with a gap condition.
\emph{Comm. Math. Phys., }\textbf{275}(1), 139--162, 2007.

\bibitem {Ni1}F. Nier. The dynamics of some quantum open systems with short
range nonlinearities. \emph{Nonlinearity}, \textbf{11}(4), 1127-1172, 1998.

\bibitem {Ni2}F. Nier. Accurate WKB approximation for a 1D problem with low
regularity. Serdica Math. J., \textbf{34} no1, 113--126, 2008.

\bibitem {PrSj}C. Presilla, J. Sj\"{o}strand. Transport properties in resonant
tunneling heterostructures. \emph{J. Math. Phys., }\textbf{37}(10), 4816-4844, 1996.

\bibitem {PrSj1}C. Presilla, J. Sj\"{o}strand. Nonlinear resonant tunneling in
systems coupled to quantum reservoirs. Phys. Rev. B: Condensed matter,
\textbf{55} no15, 9310-9313, 1997
\end{thebibliography}
\end{document}